\documentclass{elsart}

\usepackage{graphicx,latexsym,amssymb,epsf,subfigure,cite}

\newcommand{\be}{\begin{equation}}
\newcommand{\ee}{\end{equation}}

\newcommand{\bfn}{{\bf n}}
\newcommand{\bfp}{{\bf p}}
\newcommand{\bfrho}{{\bf \rho}}
\newcommand{\bfE}{{\bf E}}
\newcommand{\bfJ}{{\bf J}}
\newcommand{\bfP}{{\bf P}}
\newcommand{\bfsigma}{{\mbox{\boldmath $\sigma$}}}
\newcommand{\bfSigma}{{\mbox{\boldmath $\Sigma$}}}

\begin{document}

\begin{frontmatter}
\title{\bf Simulations of amphiphilic fluids using mesoscale 
lattice-Boltzmann and lattice-gas methods}

\author[qmul]{P. J. Love}
\author[BT]{M. Nekovee}
\author[qmul]{P. V. Coveney}
\author[qmul]{J. Chin}
\author[qmul]{N. Gonz{\'a}lez-Segredo}
\author[oxagen]{J. M. R. Martin}
\address[qmul]{Centre for Computational Science, Queen Mary, University of London, Mile End Road, London E1 4NS, U.K.}
\address[BT]{BTExact, Adastral Park, Martlesham, Ipswich, IP5 3RE, UK}

\address[oxagen]{Oxagen Limited, 91 Milton Park, Abingdon, Oxfordshire, OX14 4RY, UK} 

\begin{abstract}
We compare two recently developed mesoscale models of binary immiscible and ternary amphiphilic fluids. We describe and compare the algorithms in detail and discuss their stability properties. The simulation results for the cases of self-assembly of ternary droplet phases and binary water-amphiphile sponge phases are compared and discussed. Both models require parallel implementation and deployment on large scale parallel computing resources in order to achieve reasonable simulation times for three-dimensional models. The parallelisation strategies and performance on two distinct parallel architectures are compared and discussed. Large scale three dimensional simulations of multiphase fluids requires the extensive use of high performance visualisation techniques in order to enable the large quantities of complex data to be interpreted. We report on our experiences with two commercial visualisation products: AVS and VTK. We also discuss the application and use of novel computational steering techniques for the more efficient utilisation of high performance computing resources.  We close the paper with some suggestions for the future development of both models.
\end{abstract}



\end{frontmatter}

\section{Introduction}

The lattice-gas automata (LGA) \cite{bib:review1_lga, bib:fchc} and the lattice-Boltzmann (LBE) methods \cite{bib:lb_review1, bib:lb_review2,bib:lb_review3, bib:lb_review4} are relatively new approaches in computational fluid dynamics. Both methods have a mesoscopic character, as opposed to the conventional continuum approach based on numerical solution of the Navier-Stokes equations and the microscopic approach based on molecular dynamics. 
The key idea behind LGA is to  model fluid flows by simplified kinetic equations which describe time-evolution of particles having a discrete set of velocities while moving on a regular lattice. For the purpose of computer efficiency an ``exclusion principle'' is imposed such that at a given time step only one particle with a given velocity can occupy any lattice site. The computationally demanding tracking of individual molecules is thus avoided and at the same time macroscopic or hydrodynamic effects naturally emerge from mesoscale lattice-gas dynamics, provided that the LGA collision operator possesses the correct and necessary conservation laws and the underlying lattice has sufficient symmetry \cite{bib:review1_lga, bib:fchc}. 

Historically, the lattice-Boltzmann model was developed from lattice-gas automata by replacing the particle occupation numbers (Boolean variables) by single-particle distribution functions (real variables). Furthermore, the  LGA collision operator, which is a stochastic many-body collision operator, is massively simplified in LBE methods using the assumption of molecular chaos. Usually a simple BGK approximation~\cite{bib:bgk} is also employed, which is linear and deterministic, although more complex linearised collision operators may also be used~\cite{bib:lb_review1, bib:lb_review2,bib:lb_review3, bib:lb_review4,bib:luo2}. The spontaneous  fluctuations of LGA are thus eliminated in LBE. An important feature of LGA and LBE is the locality of the updating rules which  make these methods ideal for parallel processing.

Both methods have found applications in fluid dynamical problems in which more conventional continuum approaches face difficulties, such as multiphase and multicomponent flow and flow in porous media. LBE has also been applied to study colloidal suspensions \cite{bib:lb_review3,bib:lb_review4}, reaction-diffusion systems and other complex flow situations \cite{bib:lb_review1, bib:lb_review2,bib:lb_review3, bib:lb_review4}. An important area of application of these methods is for modelling amphiphilic fluids which consist of two immiscible phases (such as oil and water), together with an amphiphile (surfactant) species, such as detergent \cite{bib:gelbart,bib:gs2}. The study of these fluids is of relevance to a wide variety of industrial, chemical and biological applications and is of great fundamental interest. There is, however, no general consensus on what continuum hydrodynamic equations are most appropriate for describing such complex fluids. Moreover, due to the presence of complicated interfaces which undergo topological change, continuum methods based on numerical solution of Navier-Stokes-type equations face great difficulties in dealing with such fluid mixtures. Atomistic approaches based on molecular dynamics which deal with  ``real'' surfactant molecules are able to deal with such systems but are still computationally too demanding to access large time dynamics  involved in many problems of interest, such as self-assembly of micelles and the formation of lamellar and other ordered mesophases~\cite{bib:mdmlc}.

In the last few years we have pioneered the development of mesoscale lattice-gas \cite{bib:bce,bib:bcp, bib:molsim, bib:LCB, bib:LCB2, bib:LCB3} and, more recently, lattice-Boltzmann  methods \cite{bib:maziar1,bib:maziar2} for the simulation of the non-equilibrium hydrodynamic properties of amphiphilic fluids. The common features of both models is that they are ternary vector models which treat amphiphilic molecules as a separate species with an orientational degree of freedom, and implement coupling between fluid components from the ``bottom-up'' by introducing interactions between particles. Our LGA studies of amphiphilic fluids, both in two \cite{bib:bce,bib:em1} and three dimensions \cite{bib:bcp, bib:molsim, bib:LCB, bib:LCB2, bib:LCB3,bib:LCB}, demonstrate the ability of the LGA scheme to describe the non-equilibrium dynamics and hydrodynamics of these systems while our recent LBE studies in two-dimensions \cite{bib:maziar1,bib:maziar2} show that certain complex fluids phenomenologies, such as microemulsion states can be produced by both schemes. At the same time, these studies have highlighted striking differences in the phenomenology accessible to each method \cite{bib:maziar1,bib:maziar2}.

The aim of the present article is to provide a comparison between our lattice-gas and lattice-Boltzmann methods for amphiphilic fluids, in order to elucidate the individual strengths and weaknesses of these schemes, in terms of their capability to reproduce in a consistent way the phenomenology of amphiphilic systems and from the point of view of their numerical stability, computational efficiency and speed. We also describe in detail the unified parallelization strategy which we have applied to both algorithms and present the results of extensive benchmark studies of the parallel performance efficiency and speed of the resulting codes, performed on two completely different massively parallel platforms.

The rest of this paper is organised as follows. In Section~\ref{models} we provide a short overview of our lattice-gas and lattice-Boltzmann algorithms for amphiphilic fluids and discuss the relation between the two schemes and their similarities and differences. Particular emphasis will be given to the differences in the treatment of intermolecular collisions and their consequences for algorithmic complexity of the methods. In order to be clear in our discussions of both implementational and theoretical issues, we shall refer to the lattice-Boltzmann and lattice gas methods by the initials LGA and LBE, respectively, and to our particular implementations of them by the initials ME3D for our lattice gas code and LB3D for our lattice-Boltzmann code. In Section~\ref{parallel} we describe in detail our parallelization strategy and present results of our benchmark studies of the parallel performance of the LGA and LBE codes on CRAY T3E and Origin 2000 parallel supercomputers. In Section~\ref{steering} we discuss the high performance visualisation techniques ubiquitous in the understanding of results from both models. We describe our implementation of computational steering for ME3D and LB3D on parallel platforms and discuss our experience with the method. We close this paper in Section~\ref{concs} with some conclusions and outlook for the future development of the schemes.

\section{Amphiphilic lattice-gas and lattice-Boltzmann dynamics}\label{models}

In this section, we first summarise our lattice gas and lattice-Boltzmann models, and then make a specific algorithmic comparison between them.

\subsection{Amphiphilic lattice-gas fluids}
Our lattice-gas model \cite{bib:bce,bib:bcp, bib:molsim, bib:LCB, bib:LCB2, bib:LCB3} is based on a microscopic, bottom-up approach, where dipolar amphipile particles are included alongside the immiscible oil and water species. The model was developed as an extension of the Rothman and Keller (RK) \cite{bib:rk} model of binary immiscible fluids, including its generalization by Chan and Liang \cite{bib:chli}, to take  into account the essential characteristics and interactions of amphiphiles. In the RK model \cite{bib:rk} the lattice-gas particles are tagged with two ``colours'', to distinguish oil and water. The immiscible fluid effect is incorporated by modifying the LGA collision operator such that the collisions favor outcomes that send particles towards neighbouring sites dominated by other particles of the same colour (under the constraint that mass of of each species and the total momentum of both species together is locally conserved). This  is achieved by sampling the collision outcome from the Gibbsian equilibrium corresponding to a local, sitewise Hamiltonian function which depends on the net colour charge associated with a given site 
\be 
q_i{\left({\bf x},t\right)}\equiv n^R_i{\left({\bf x},t\right)}-n^B_i{\left({\bf x},t\right)}, 
\ee 
and the colour flux vector of an outgoing state
\be
\bfJ{\left({\bf x},t\right)}\equiv\sum_{i=1}^b{\bf c}_i q'_i{\left({\bf x},t\right)},
\label{eq:cflux} 
\ee
where the prime denotes a post-collision quantity.
 
The amphiphilic model is obtained by introducing surfactant particles in the RK lattice gas as a third species. In nature, amphiphilic molecules usually possesses two different fragments, each having an affinity for one of the two immiscible components \cite{bib:gs2}. In the case of a prototype mixture of immiscible fluids like oil and water, a typical surfactant molecule would be an amphiphile that has a hydrophilic head preferring to be in contact with water molecules, and a hydrophobic tail preferring to be surrounded by oil. This crucial property is taken into account in the lattice gas model by assigning to each surfactant particle at position ${\bf x}$ moving in the direction $i$ a ``colour dipole vector'' ${\bf \sigma}_i({\bf x},t)$ which models the orientational degree of freedom of these molecules. In analogy with electrostatics the colour Hamiltonian of the RK model is then extended to take into account dipolar interactions among surfactant molecules themselves and between surfactant particle and colour charges \cite{bib:bce}.

Lattice-gas particles have velocities ${\bf c}_i$, where $1\leq i\leq b$, and $b$ is the number of velocities per site. A particle emerging from a collision at site ${\bf x}$ and time $t$ with velocity ${\bf c}_i$ streams along its velocity vector to site ${\bf x}+{\bf c}_i\Delta t$ where it may undergo the next collision. We let $n^\alpha_i{\left({\bf x},t\right)}\in\{0,1\}$ denote the presence ($1$) or absence ($0$) of a particle of species $\alpha\in\{R,B,A\}$ ($R$, $B$, $A$ denoting red (oil), blue (water) and green (amphiphile) species respectively) with velocity ${\bf c}_i$, at lattice site ${\bf x}$ and time step $t$. The collection of all $n^\alpha_i{\left({\bf x},t\right)}$ for $1\leq i\leq b$ will be called the {\it population state} of the site; it is denoted by ${\bf n}{\left({\bf x},t\right)}$. The dipole vector of amphiphilic particles ${\bf \sigma}_i$ is assumed to have a fixed magnitude but its orientation can vary continuously. The collection of the $b$ vectors $\bfsigma_i{\left({\bf x},t\right)}$ at a given site ${\bf x}$ and time step $t$ is called the {\it orientation state}.  

The time evolution of the system is an alternation between an streaming or {\it propagation} step and a {\it collision step}. The propagation step is described mathematically by the replacements
 \be
n^\alpha_i\left({\bf x}+{\bf c}_i,t+1\right) 
\leftarrow n^\alpha_i{\left({\bf x},t\right)}, 
\label{eq:propp} 
\ee

\be
\bfsigma_i\left({\bf x}+{\bf c}_i,t+1\right) \leftarrow
\bfsigma_i{\left({\bf x},t\right)}, \label{eq:propo} \ee 
for all ${\bf x}$, $1\leq i\leq b$ and $\alpha\in\{ R,B,A\}$.  That is, particles with velocity ${\bf c}_i$ simply move from point ${\bf x}$ to point ${\bf x}+{\bf c}_i$ in one time step. In the collision step, the newly arrived particles interact, resulting in new momenta and surfactant orientations. The collisional change in the state at a lattice site ${\bf x}$ is required to conserve the mass of each species present 
\begin{equation}
\rho^\alpha{\left({\bf x},t\right)}\equiv\sum_i^b n^\alpha_i{\left({\bf x},t\right)}, 
\end{equation} as well as the $D$-dimensional momentum vector 
\begin{equation}
\bfp{\left({\bf x},t\right)}\equiv\sum_\alpha\sum_i^b{\bf c}_i n^\alpha_i{\left({\bf x},t\right)}, 
\end{equation}
(where we have assumed for simplicity that the particles all carry unit mass).  Thus, the set of population states at each site is partitioned into {\it equivalence classes} of population states having the same values of these conserved quantities. 

The  Hamiltonian for the amphiphilic model is much more complex than the original RK Hamiltonian and is derived and described in detail in \cite{bib:bce}. It is given by
\begin{equation}
H(s') = \bfJ\cdot (\alpha\bfE+\mu\bfP) + \bfsigma'\cdot
(\epsilon\bfE+\zeta\bfP) + {\bf Q} : (\epsilon{\bf R}+\zeta{\bf M})+{\delta
\over 2} {{{\bf v}({\bf x},t)}^{2}}, \label{eq:hamil} 
\end{equation} 
where the various quantities in this expression are defined in Table~\ref{tab:one}
\begin{table}[h]
\caption{\label{tab:one}Quantities required for LGA collision step}
\begin{tabular}{|l|l|}
\hline
{\bf Quantity}& {\bf Mathematical expression}\\
\hline
Total director at a site & $\bfsigma{\left({\bf x},t\right)}\equiv\sum_{i=1}^b\bfsigma_i{\left({\bf x},t\right)}$ \\
\hline
Dipolar flux tensor of an outgoing state & ${\bf Q}{\left({\bf x},t\right)}\equiv\sum_{i=1}^b{\bf c}_i\bfsigma'_i{\left({\bf x},t\right)}$\\
\hline
Colour field vector & $\bfE{\left({\bf x},t\right)}\equiv\sum_{i=1}^b{\bf c}_i q{\left({\bf x}+{\bf c}_i,t\right)}$ \\
\hline
Dipolar field vector & $\bfP{\left({\bf x},t\right)}\equiv-\sum_{i=1}^b{\bf c}_i S{\left({\bf x}+{\bf c}_i,t\right)}$ \\
\hline
Colour field gradient tensor & ${\bf R}{\left({\bf x},t\right)}\equiv\sum_{i=1}^b{\bf c}_i\bfE{\left({\bf x}+{\bf c}_i,t\right)}$ \\
\hline
Dipolar field gradient tensor & ${\bf M}{\left({\bf x},t\right)}\equiv-\sum_{i=1}^b{\bf c}_i{\bf c}_i S{\left({\bf x}+{\bf c}_i,t\right)}$\\
\hline
Scalar director field & $S{\left({\bf x},t\right)}\equiv\sum_{i=1}^b{\bf c}_i\cdot\bfsigma_i{\left({\bf x},t\right)}$\\
\hline
Kinetic energy & ${\delta \over 2}\left|\frac{(\sum_{i=1}^b n_i^\sigma{\bf c}_i({{\bf x}},t)}{\sum_\alpha \rho_\alpha }\right|^2$\\
\hline
\end{tabular}
\end{table}\\
The mass of the particles is taken as unity, and $\alpha$, $\mu$, $\epsilon$, $\zeta$ and $\delta$ are coupling constants.

In the above equations the term parameterised by $\alpha$ models the
interaction of colour charges with surrounding colour charges as in the
original Rothman-Keller model~\cite{bib:rk}; that parameterised by $\mu$
describes the interaction of colour charges with surrounding colour
dipoles; that parameterised by $\epsilon$ accounts for the interaction of
colour dipoles with surrounding colour charges (alignment of surfactant
molecules across oil-water interfaces); and finally that parameterised by
$\zeta$ describes the interaction of colour dipoles with surrounding
colour dipoles (corresponding to interfacial bending energy or ``stiffness''). 
The outgoing states of the collisions are sampled from 
\be 
P(s') = \frac{1}{Z}\exp\left[-\beta H(s')\right],
\label{eq:beta_defn} 
\ee 
where $\beta$ is an inverse
temperature, $H(s')$ is the energy associated with collision outcome $s'$,
and $Z$ is the equivalence-class partition function, 
\begin{equation} Z
(\bfrho,\bfp,\beta) \equiv \sum_{\bfn'\in E(\bfrho,\bfp)} \int
d\bfSigma'\;\exp\left[-\beta H(s')\right], 
\end{equation} 
and we have defined the measure on the set of orientational states 
\begin{equation}
\int d\bfSigma\equiv\prod_{i=1}^b\int d\bfsigma_i.
\end{equation} 
In practice, we sample $s'=\left(\bfn',\bfSigma'\right)$ by first sampling
the postcollision population state $\bfn'$ from the reduced probability
density 
\be 
P\left(\bfn'\right) = \int d\bfSigma'\;{\bf M}(s'). 
\label{eq:redp} \ee
We then sample the postcollision orientation state by sampling the $b$ orientations $\bfsigma_i' $ from
each of 
\be
\pi_i\left(\bfsigma'_i\right) = \prod_{j\neq
i}^b\int d\bfsigma'_j\;{\bf M}\left(\bfn',\bfSigma'\right). 
\label{eq:redo}
\ee{equation} 
for $1\leq i\leq b$; these are
independent distributions, so the $b$ samples may each be taken without
regard for the other outcomes.  

In the current implementation of the above algorithm in 3D we use 
a projected face-centered hypercubic (PFCHC) lattice \cite{bib:fchc}.
The motivation for using this lattice is that it is known to yield 
isotropic Navier-Stokes behavior for a single-phase fluid \cite{bib:fchc}. 

\subsection{Amphiphilic lattice-Boltzmann fluids}
In close analogy with the above amphiphilic lattice-gas model, our amphiphilic lattice-Boltzmann scheme is obtained by introducing surfactant molecules as a third species in the immiscible two component lattice-Boltzmann scheme of Shan and Chen \cite{bib:shanchen1, bib:shanchen2}. In this  scheme \cite{bib:shanchen1, bib:shanchen2}, coupling between fluid components is achieved by introducing pair-wise interactions between particles, within  a mean field approximation. The resulting force acting on particles of component $\sigma$ is modelled as 
\be
{\bf F}^{\sigma \bar \sigma}(\bf x,t) = -\psi^\sigma (\bf x,t)  \sum_{\bar \sigma} 
g_{\sigma \bar \sigma} \sum_{{\bf x}'}  G_{\sigma 
{\bar \sigma}}({\bf x},{\bf x}') \psi^{\bar \sigma} ({\bf x}',t) ({\bf x}' -  {\bf x}).
\label{eq:force}
\ee
Here $G_{\sigma \bar \sigma}({\bf x}, \bf x')$ is an interaction
kernel and $\psi^\sigma=\psi^\sigma(n^{\sigma}({\bf x},t))$ is a 
function of density, whose  form is chosen empirically
to model various types of fluids. Finally, 
$g_{\sigma {\bar \sigma}}$ ($>0$ for immiscible fluids) is 
a  coupling constant, whose magnitude controls interfacial tension of the
binary immiscible fluid \cite{bib:shanchen1, bib:shanchen2}.

In the original Shan-Chen model the above force term is incorporated in an {\it ad hoc} fashion by adding an increment \cite{bib:shanchen1, bib:shanchen2}
\be
\delta {\bf u}^{\sigma} = 
\frac{{\bf F}^{\sigma}}{\rho^\sigma} \Delta t
\ee
to the velocity $\tilde{{\bf u}}$ which enters the equilibrium distribution function in the BGK collision operator. A more rigorous inclusion of the force term in LBE models, however, can be achieved  by discretization of the continuum Boltzmann equation in the presence of a body force \cite{bib:luo1,bib:martys}. An advantage of this approach is that, unlike the original Shan-Chen model, the force term does not enter in the equilibrium distribution function, hence making a Chapman-Enskog expansion of the LBE equations straightforward. Furthermore, it can be shown that with the choice $\psi^{\sigma}=n_{\sigma}$ the phenomenological Shan-Chen model maps exactly onto  a discretised version of the Boltzmann-Vlasov equations of binary interacting fluids \cite{bib:basleb,bib:basleb2,bib:maziar2}.

Following the treatment of surfactant molecules in the amphiphilic LGA scheme the surfactant molecules are assumed to carry a dipole vector (or ``director''). Instead of specifying directors for each surfactant particle, however, this degree of freedom is further coarse-grained by introducing a dipole vector field ${\bf d}({\bf x},t)$ which represents the average director  of all amphiphilic particles present at site ${\bf x}$ at time $t$. Once again, the orientation of ${\bf d}$ is allowed to vary continuously in time but its magnitude is a fixed input of the calculations. Incorporation of dipole-carrying amphiphilic fluid particles results in two fundamentally new types of interaction forces (between amphiphilic and non-amphipilic particles and among amphiphiles themselves) which depend not only on the relative distance between particles but also on the dipolar orientations. These forces are derived from~(\ref{eq:force} by treating each amphiphilic molecule as a pair of water and oil molecules displaced by a distance ${\bf d}({\bf x},t)$ from each other and performing a Taylor expansion in ${\bf d}$ in the resulting expression for the total force
\cite{bib:maziar1}.
Assuming that the dipole head and tail have equal and opposite colours $e=\pm1$ and only nearest-neighbour interactions considered, the additional forces are given by \cite{bib:maziar1}
\begin{equation}\label{bscn}
{\bf F}^{\sigma , s}({\bf x},t)= - 2\psi^\sigma ({\bf x},t) g_{\sigma s}\sum_{i\neq 0}{\bf d}({\bf x} + {\bf c}_i  \Delta t)\cdot ({\bf I} - \frac {{\bf c}_i {\bf c}_i} {c_i^2} D)\psi^s ({\bf x} + {\bf c}_i  \Delta t,t)
\end{equation}
\begin{equation}\label{bcsn}
{\bf F}^{s, c}({\bf x},t)= 2\psi^s ({\bf x},t) {\bf d}({\bf x},t)\cdot \sum_\sigma g_{\sigma s}\sum_{i \neq 0} ({\bf I} - \frac {{\bf c}_i {\bf c}_i} {c_i^2} D)\psi^\sigma ({\bf x} + {\bf c}_i,t)
\end{equation}
and 
\begin{eqnarray}
{\bf F}^{s,s}({\bf x},t) &=& - \frac {4D} {c^2}
g_{ss} \psi^s({\bf x}) \sum_i
{\bf d}({\bf x} + {\bf c}_i \Delta t,t) {\bf d}({\bf x},t) :
[{\bf I} - \frac {{\bf c}_i {\bf c}_i} {c_i^2} D] {\bf c}_i \nonumber \\
&+&   {\bf d}({\bf x} + {\bf c}_i \Delta t,t) {\bf d}({\bf x},t)\cdot {\bf c}_i  \nonumber \\
&+& {\bf d}({\bf x},t) {\bf d}({\bf x} + {\bf c}_i  \Delta t,t) \cdot {\bf c}_i \} \psi^s({\bf x} + {\bf c}_i + \Delta t,t)\nonumber \\
\label{eq:nssf}
\end{eqnarray}
In the above equations  ${\bf I}$ is the second-rank unit tensor,
${\bf F}^{\sigma,s}$ is the force acting on 
non-amphiphilic particles $\sigma$ (water and oil) 
due to amphiphile dipoles, ${\bf F}^{s,c}$ is the force 
acting on amphiphilic particles due to all non-amphiphilic particles and 
${\bf F}^{s,s}$ is the force among amphiphilic particles themselves. 
The  coupling constants $g_{\sigma s}$ and $g_{s s}$ determine,
respectively, the strength of interaction between water/oil particles
and surfactant particles, and among surfactant particles themselves.

The resulting amphiphilic lattice-Boltzmann model is characterised by the following set of coupled equations \cite{bib:maziar2}:
\begin{eqnarray}\label{eq:lba1}
f^{\sigma}_i({\bf x} + {\bf  c}_i\Delta t, t + \Delta t) - 
f^{\sigma}_i({\bf x}, t) &=& 
-\Delta t \frac {f^\sigma_i-f_i^{\sigma (eq)} }{\lambda_\sigma}\nonumber \\
&+& \sum_{{\bar \sigma}}\sum_j \Lambda_{ij}^{\sigma {\bar{\sigma}}} f_j^{{\bar{\sigma}}}\nonumber \\
&+& \sum_j \Lambda_{ij}^{\sigma s} f_j^{s}
\end{eqnarray}
\begin{eqnarray}\label{eq:lba2}
f^{s}_i({\bf x} + {\bf  c}_i\Delta t, t + \Delta t) - 
f^{s}_i({\bf x}, t) &=& -\Delta t \frac {f^s_i-f_i^{s (eq)} }{\lambda_s}\nonumber \\
&+& \sum_{\sigma}\sum_j \Lambda_{ij}^{s \sigma} f_j^{\sigma} \nonumber \\
&+& \sum_j \Lambda_{ij}^{s s} f_j^{s}\nonumber \\
\end{eqnarray}
\be
{\bf d}({\bf x},t+\Delta t) - \bar{{\bf d}}({\bf x},t) = -\Delta t
\frac{\bar{{\bf d}}({\bf x},t)-{\bf d}^{(eq)}({\bf x},t)}{\lambda_d}.
\label{eq:lba3}
\ee
The first two equations describe time evolution of discrete velocity distribution functions $f^\sigma_i$ and  $f^s_i$ belonging to component $\sigma$ ($\sigma$ oil, water) and  surfactant (s), respectively and the third equation describes time evolution of surfactant dipoles ${\bf d}(x,t)$. In the above equations $f_i^{\sigma (eq)}$, $f^{s (eq)}_i$ and ${\bf d}^{(eq)}$ are suitably chosen local equilibrium distribution functions \cite{bib:maziar1}, $\lambda_{\sigma}$, $\lambda_s$  and $\lambda_d$ are relaxation times and ${\bar d}$ is the average dipole at site ${\bf x}$ prior to collisions. The first term on the right-hand sides of Eqns.~(\ref{eq:lba1}) and~(\ref{eq:lba2}) is the standard BGK collision operator~\cite{bib:lb_review1, bib:lb_review2,bib:lb_review3, bib:lb_review4}. The terms $\Lambda_{ij}^{\sigma {\bar{\sigma}}}$, $\Lambda_{ij}^{s \sigma}$, $\Lambda_{ij}^{s \sigma}$ and $\Lambda_{ij}^{s s}$  are matrix elements of collision operators which result from mean-field interactions among different fluid components~\cite{bib:maziar2}. For Example $\Lambda_{ij}^{\sigma \bar \sigma }$ is given by:
\be
\Lambda_{ij}^{\sigma {\bar{\sigma}}} = \omega_i 
\left[\frac{1}{c_s^2}
 (\delta_{\sigma {\bar{\sigma}}}{\bf c}_i- \alpha_{\sigma {\bar{\sigma}}}
{\bf c}_j)+ \alpha_{\sigma,{\bar{\sigma}}}
\frac{{\bf c}_i.{\bf c_j}}{c_s^4} {\bf c}_i\right].
{\bf a}_\sigma \Delta t 
\ee
with 
\be
\alpha_{\sigma {\bar{\sigma}}} = \frac{n^\sigma}{n^{{\bar{\sigma}}}} \times
\frac {\frac {\rho^{{\bar{\sigma}}}}
{\tau_{{\bar{\sigma}}}} } { \sum_{{\bar{\sigma}}} \frac {\rho^{{\bar{\sigma}}}} {\tau_{{\bar{\sigma}}}}}.
\ee
Hydrodynamic quantities such as number densities  $n^{\sigma,s}$ and
velocities ${\bf u}^{\sigma,s}$ of each fluid component are 
obtained from velocity moments of the corresponding distribution functions.
The kinematic viscosities of each fluid component are controlled by the corresponding relaxation time $\lambda_{\sigma}$ and $\lambda_s$ \cite{bib:maziar1}. 
\subsection{Algorithmic comparison}
Each time step of our LGA and LBE algorithms consists of the following substeps:
streaming of particles or distribution functions to adjacent sites, computation of interaction fields, fluxes and forces, and local computation of the post-collision state at each site.

In the streaming step, the particle occupation numbers or distribution functions are updated according to~(\ref{eq:propp}). LGA particles carry their directors to the neighbouring site during this step and the average director of LBE particles is updated  using 
\be
n^s({\bf x},t){\bar {\bf d}}({\bf x},t) = \sum_{i} f_i^s({\bf x}-
{\bf c}_i \Delta t){\bf d}({\bf x}-{\bf c}_i\Delta t ,t).
\ee
Once the streaming substep is completed, the new fields 
$S$, ${\bf E}$, ${\bf P}$, and ${\bf R}$ and ${\bf P}$ (in LGA) 
and ${\bf F}^{\sigma \bar \sigma}$, ${\bf F} ^{\sigma s}$, 
${\bf F}^{s c}$, ${\bf F}^{s s}$ (in LBE) are computed 
at each site. 

The collision substep of the algorithms differs  greatly. The LBE collision step requires only the calculation of the equilibrium distribution functions $f^{\sigma, (eq)}_i$,$f^{s, (eq)}_i$  $d^{\sigma,(eq)}_i$ for each species and each velocity vector. Thus in going from a two-dimensional lattice (with typically $9$ velocity vectors) to a three-dimensional lattice (with typically $19$ to $23$ velocity directions) the computational time spent in the collision substep (per site) increases by less than a factor of three. The LGA collisions, however, are truly particulate and involve many-body collisions between all particles entering a site. To perform the LGA collision substep, a list of all possible states that has been sorted according to their equivalence class is recomputed and stored. In 3D the storage of the full list would require a total of $2^{52}$ bits of memory and is not feasible on any existing computer. A method for  shortening this list was described in \cite{bib:bcp, bib:molsim, bib:LCB, bib:LCB2, bib:LCB3} which drastically reduces the required memory to $12$ MB. The lookup table accepts the current population state ${\bf n}$ as input and returns a pointer to the initial position and the number of elements of the equivalent class $E({\bf \rho}, {\bf p})$ in the table of stored states \cite{bib:bcp, bib:molsim, bib:LCB, bib:LCB2, bib:LCB3}. Next each site loops over this set of allowed postcollision states and computes the required fluxes and fields shown in table~\ref{tab:one}. Note that unlike the field quantities, which depend only on the precollision state and are calculated once for each site, the fluxes must be calculated for each possible outgoing state. These are then used to compute the Gibbsian probability densities from Eqn. ~(\ref{eq:redp}) from which a final state ${\bf n'}$ is sampled. Once ${\bf n'}$ is known, the postcollision orientation states of surfactant particles are sampled from Eqn.~(\ref{eq:redo}).

\subsection{Numerical stability}

In conventional fluid dynamics (CFD) much attention is devoted to analysis of the numerical stability of Navier Stokes solvers. CFD algorithms typically become numerically unstable for critical values of the time step and/or grid size. Such considerations become paramount when pursuing the highest possible Reynolds number simulation, as, for example, when studying turbulence. When the FHP lattice gas was originally introduced one feature of particular interest was the {\it unconditional} numerical stability of the model. The introduction of our modifications to enable simulation of amphiphilic fluids retain this property. The additional floating point calculations required for computation of the local Hamiltonian are sufficiently straightforward that they do not introduce additional stability considerations.

On moving to the lattice-Boltzmann model, this appealing feature of the lattice gas is lost. The BGK approximation for the collision operator introduces hydrodynamic modes which may cause instabilities. In addition to these hydrodynamic modes, the force term in our discrete Boltzmann equation may become so large that it produces negative values of the distribution function. Such negative values are unphysical and cause the code to become unstable. The force term depends on the composition of the system, and so these instabilities are in principle unpredictable in nature. In practice, in extensive $2$ dimensional studies we have found that choices of the LB3D parameters which allow the code to run at all will allow the code to run for in excess of $2 \times 10^5$ time steps. Such long term stability is much better than that reported for a lattice-Boltzmann model based on free energies, which is currently believed to be unconditionally unstable (the code will become unstable at sufficiently long times for any parameter choice)~\cite{bib:catesJFM}.

\section{Non-equilibrium dynamics and self-assembly of amphiphilic fluids}

The non-equilibrium behaviour of the LGA and LBE models described above has been previously reported in the literature~\cite{bib:LCB}. In this section we review only those aspects of the models behaviour which yield informative comparisons.

Under normal temperature and pressure conditions, water and oil are immiscible. The addition of amphiphile can bring them together with the formation of a wealth of complex structures whose construction is driven by the properties of amphiphiles. The energy of the amphiphile is lowest when it can find or create surfaces between oil and water at which it can adsorb. Lyotropic phases such as hexagonal, lamellar and cubic phases occur when the concentration of surfactant, relative to that of oil and water, is high enough to arrange the interfaces into structures with long-range order \cite{bib:gs2}. If the concentration is not high enough to bring about the formation of lyotropic phases but more than enough to overcome the tendency of oil and water to phase separate, then an equilibrium fluid phase can be formed in which the water and oil are solubilised, with surfactant residing predominantly at the interfaces between these two components. Such a fluid is called a microemulsion \cite{bib:gs2}. Self-assembly also occurs in two component water-surfactant systems, where depending on the concentration of surfactant, disordered micellar aggregates and ordered phases can form~\cite{bib:gs2}.

The above-described phenomenology of amphiphilic fluids is extremely rich and the occurrence of various ordered and disordered phases depends not only on the relative concentrations of water, oil and surfactant, and on temperature, but also on specific molecular properties of surfactant, such as  size of the headgroups and ionic strength, {\it inter alia}~\cite{bib:gs2}. We do not expect, therefore, that minimal models of amphiphiles, like ours, could provide a complete description of all this phenomenology within the parameter space. 
However, we do expect that phenomena which depend critically on kinetic fluctuations, nucleation and other particle discreteness effects (such as formation
of micelles) are better described by our LGA model than by our LBE model.
On the other hand the lack of fluctuations in the LBE model means that self-assembly of structures with long-range order, such as lamellar\cite{bib:maziar1} and cubic phases~\cite{bib:maziar4}, are more likely to be found within this model. 

In order to demonstrate the above ideas we select a set of parameters for each model such that for a given concentration of oil, water and surfactant both models have as their final equilibrium state the oil-in-water droplet microemulsion phase. This phase is obtained when the concentration of oil in the system is lower than that of water and consists of finely divided spherical regions of oil, with stabilizing monolayers of surfactant surrounding them, embedded within a continuously connected water background. Starting from this common structure we then investigate the phase behavior of each model when the concentration of oil is set equal to zero. Both our LGA and LBE simulations  reported below are performed on a $64^{3}$ PFCHC lattice with periodic boundary conditions.

\subsection{Lattice-Boltzmann simulations}
\begin{figure}[htp]
\centering
\includegraphics[width=0.4\textwidth]{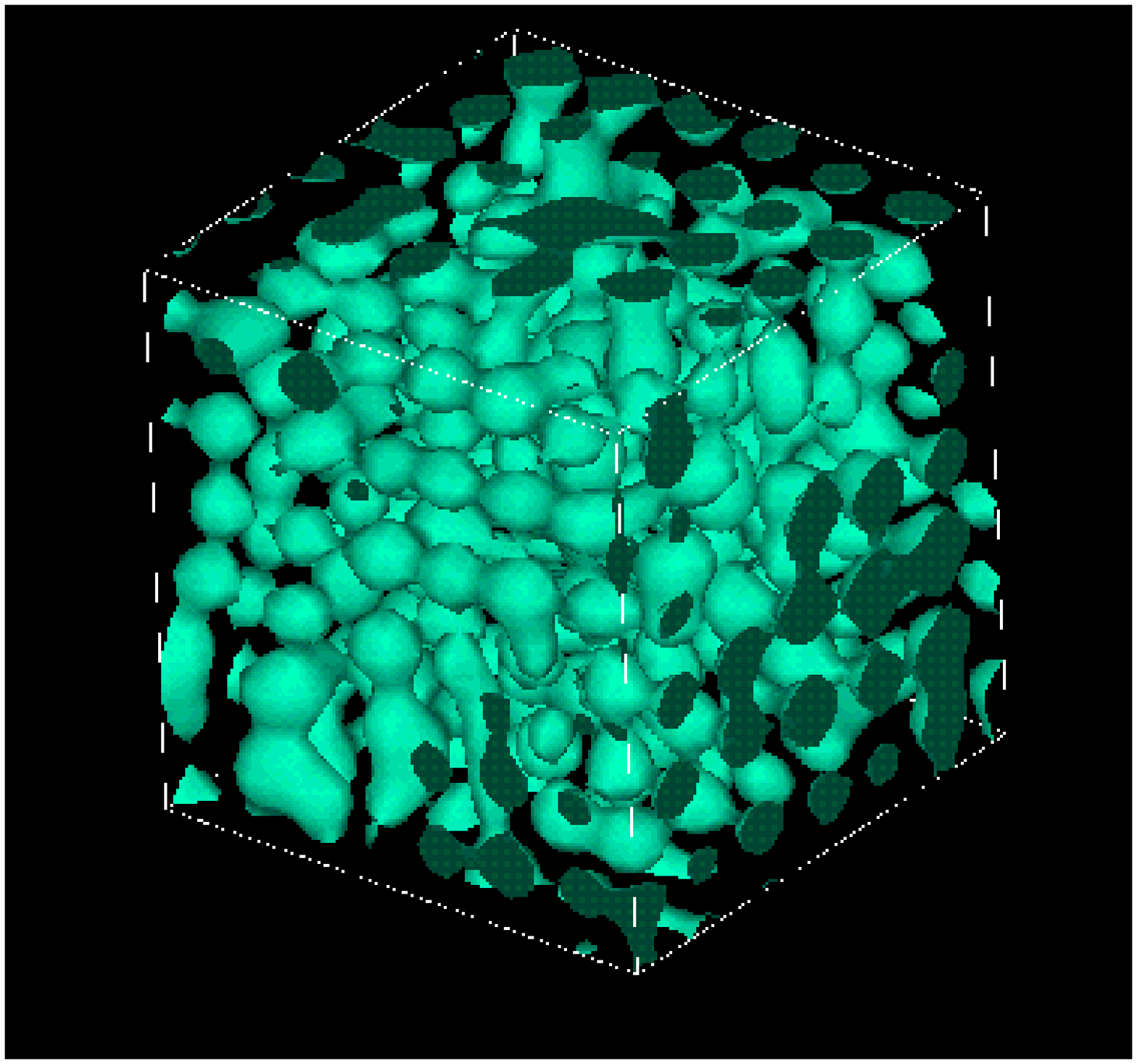}
\includegraphics[width=0.4\textwidth]{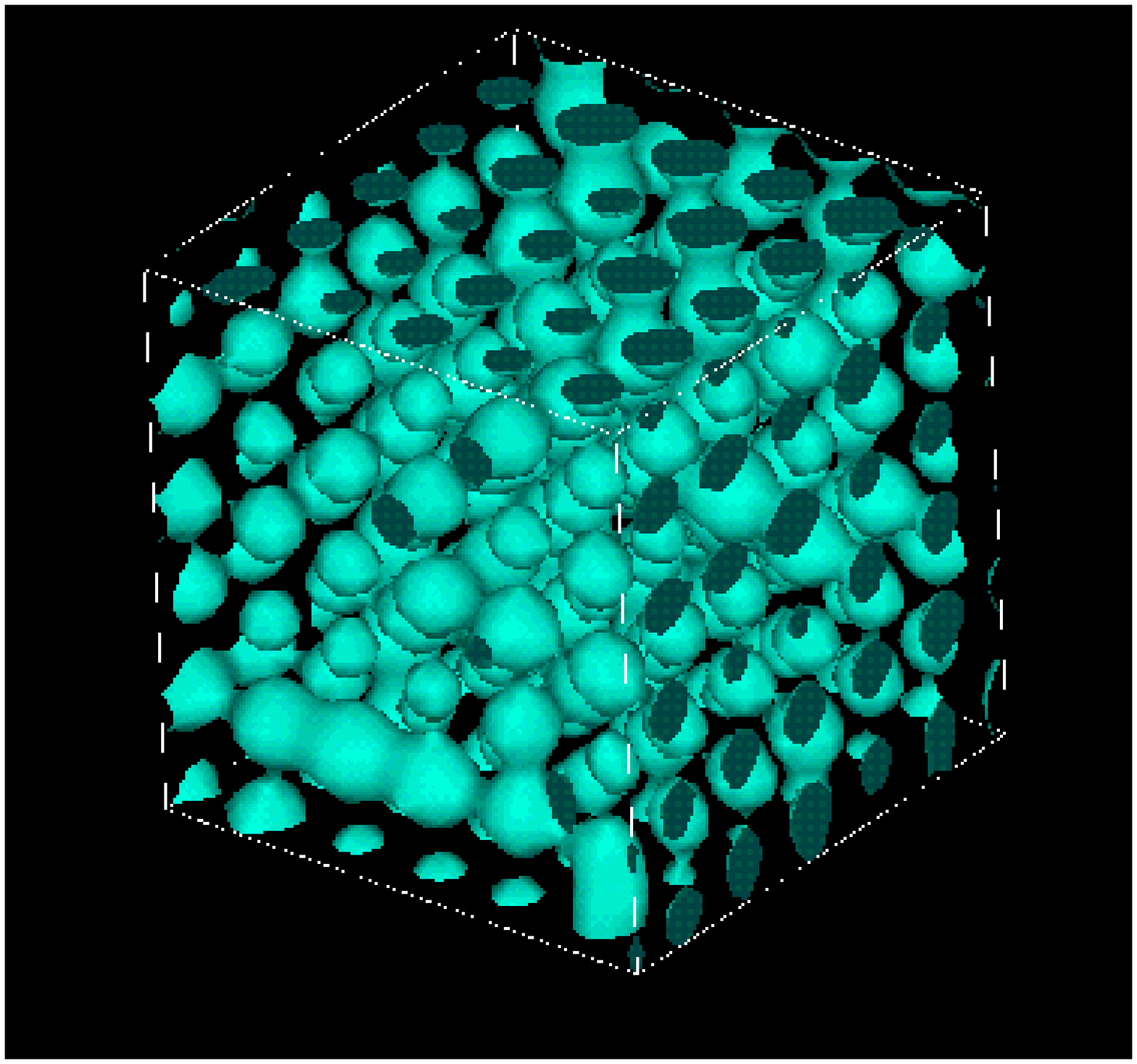}
\includegraphics[width=0.4\textwidth]{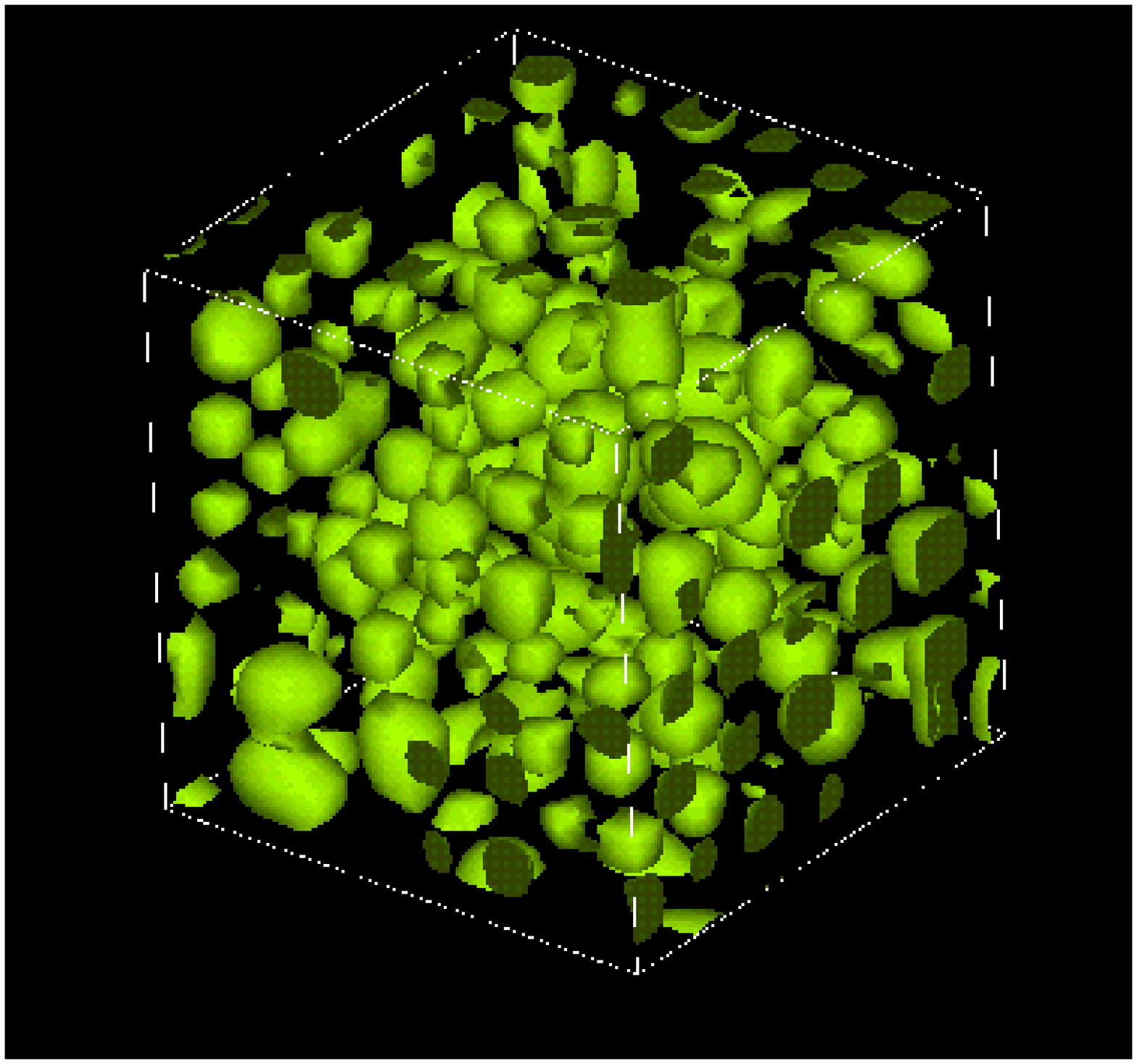}
\includegraphics[width=0.4\textwidth]{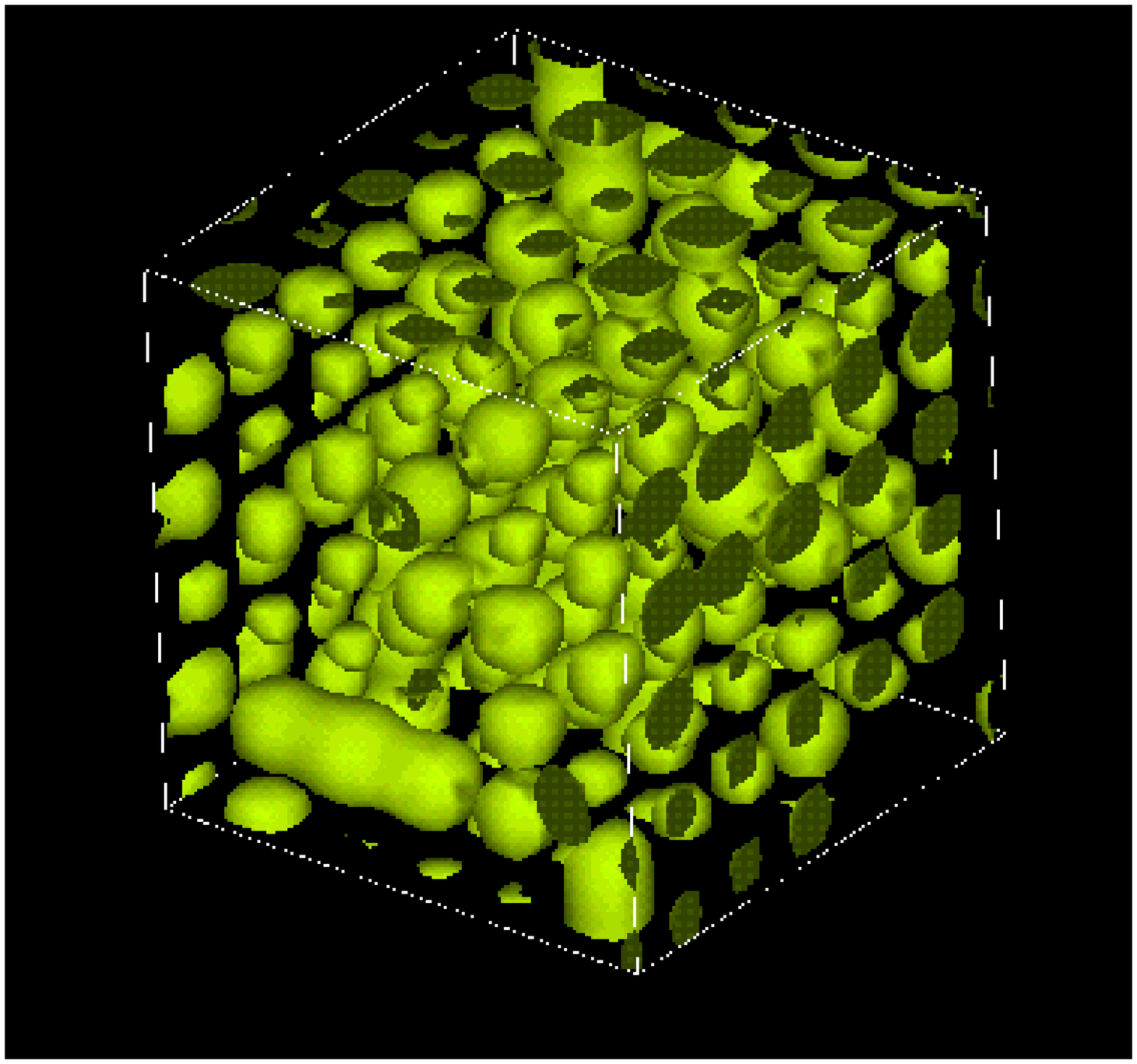}
\caption{\label{fig:one} Lattice-Boltzmann simulation of a stable ternary oil-in-water droplet phase arising from a homogenised initial condition. Upper panels shows (from left to right) the  $0.2$ oil isosurfaces at time steps $400$ and $6000$. Lower panels show the $0.27$ surfactant isosurfaces at the same time steps.}
\end{figure}

\begin{figure}[htp]
\centering
\includegraphics[width=0.4\textwidth]{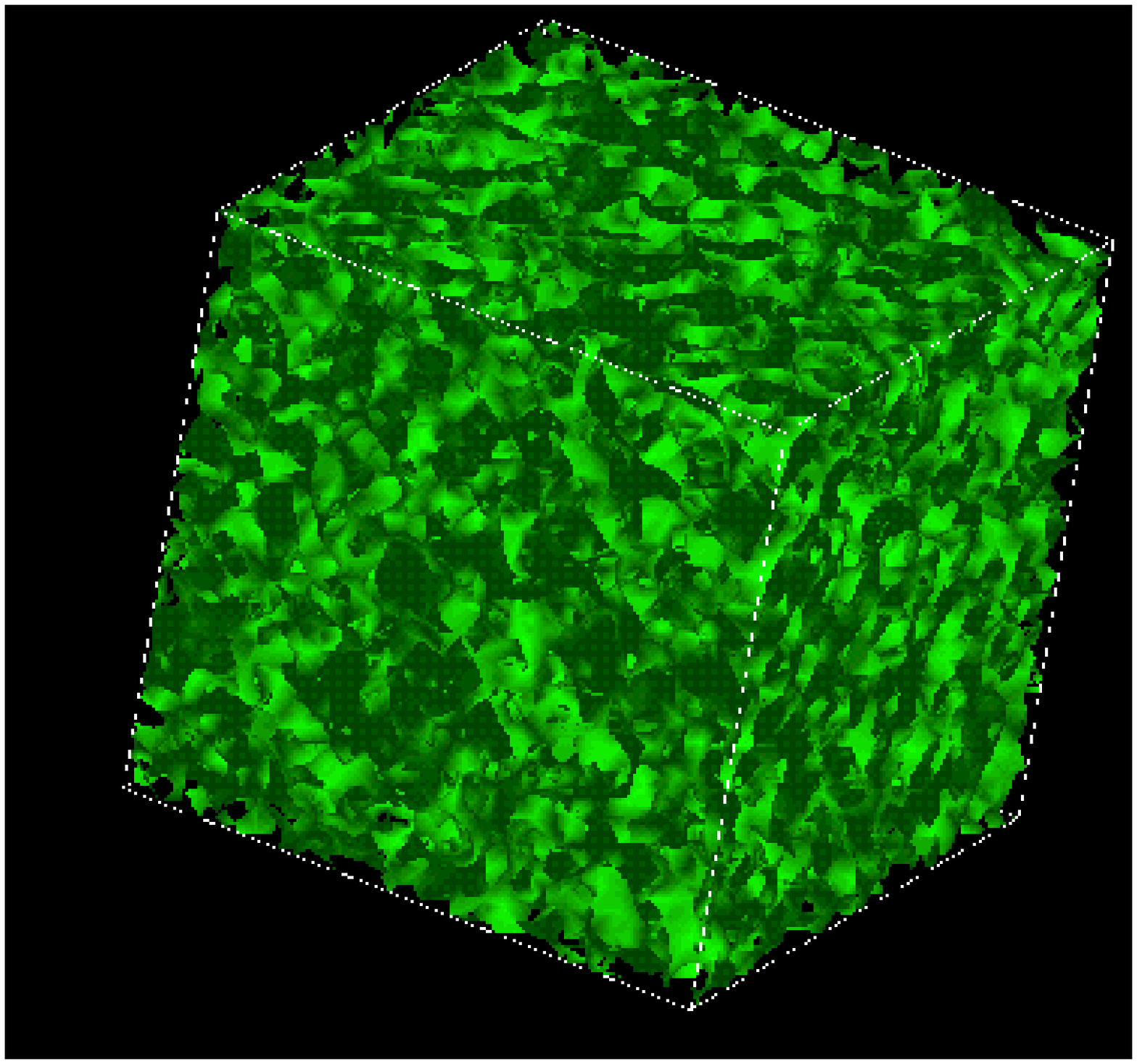}
\includegraphics[width=0.4\textwidth]{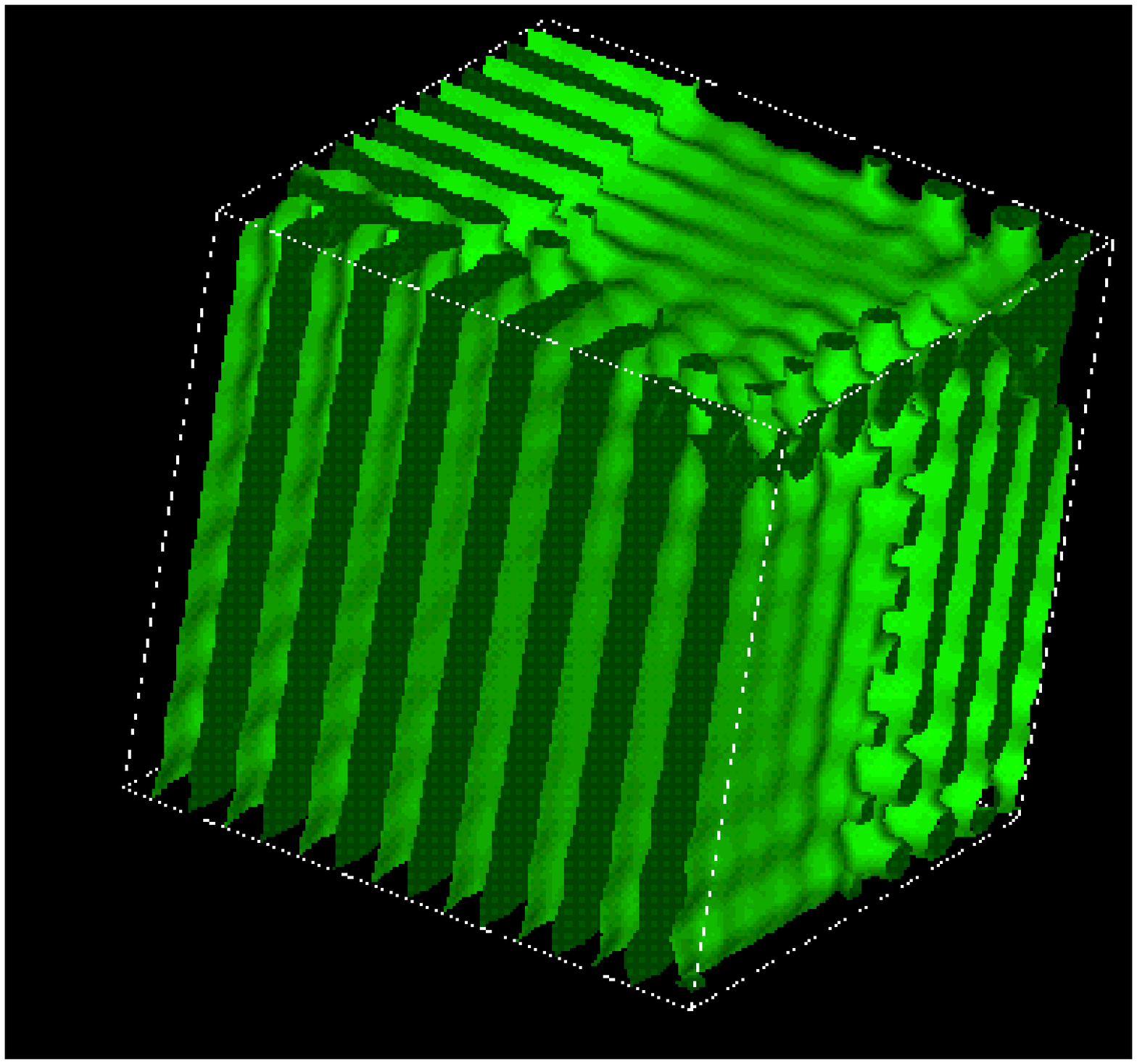}
\includegraphics[width=0.4\textwidth]{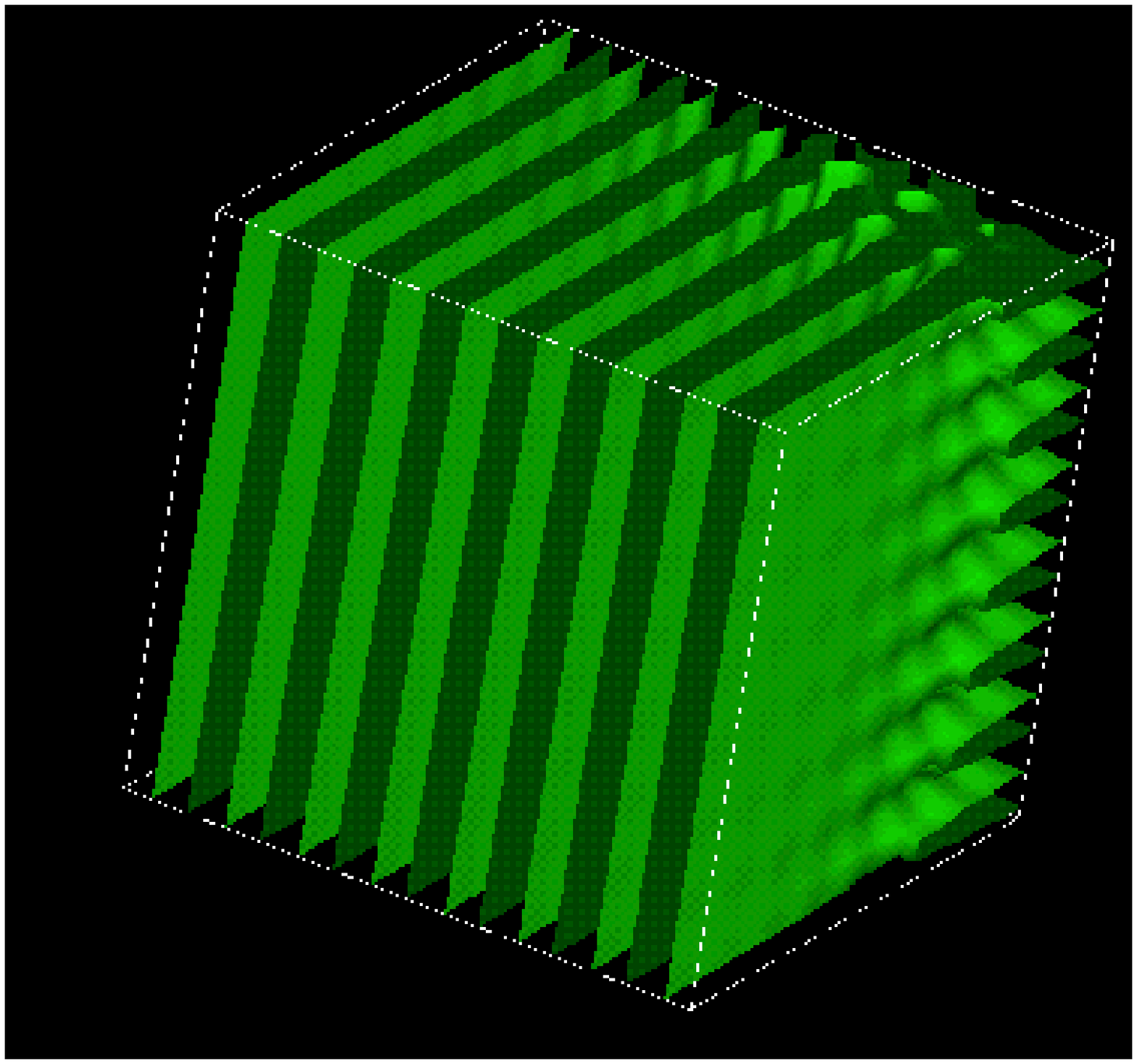}
\includegraphics[width=0.4\textwidth]{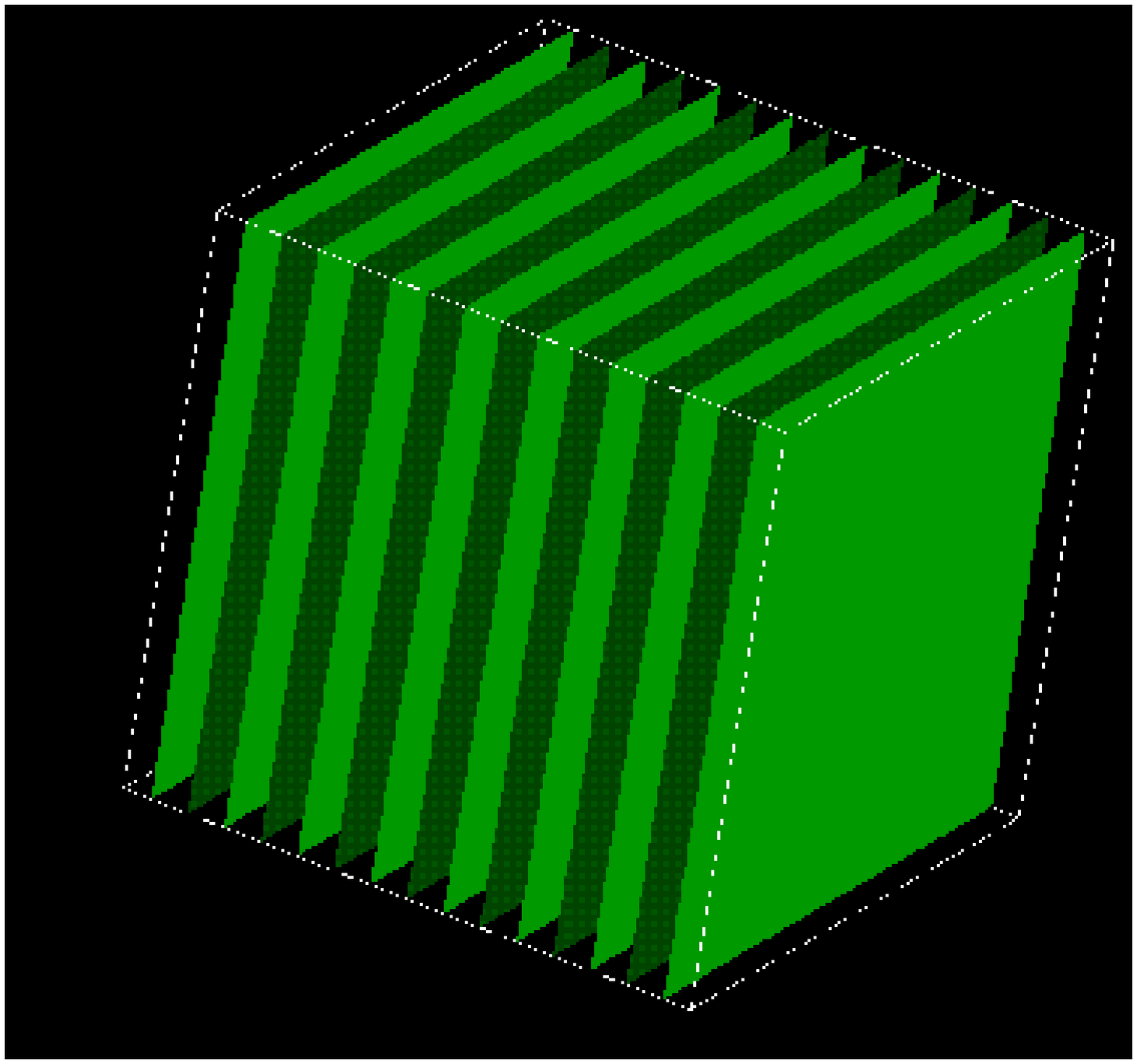}
 \caption{\label{fig:two}Lattice-Boltzmann simulation of self-assembly of the lamellar structure from a homogenised mixture of water and surfactant. From top to bottom and left to right the  surfactant midsurfaces are shown at time steps $0$, $400$, $2000$ and $3600$ of the simulations.}
\end{figure}

We used the following set of canonical parameters throughout our LBE simulations (the time step  $\Delta t$ is set to $1$ throughout) $g_{\sigma {\bar{\sigma}}}=0.05$, $g_{\sigma s} =-0.01$, $g_{ss} = 0.01$, $\tau_{\sigma}=\tau_s=1$, $\tau_d=2$, $m^{\sigma}=1$, $m^{s}=1$. 
In Figure~\ref{fig:one} we show the initial and final states of a simulation in which the average concentrations of water, oil and surfactant are set at $0.5$, $0.25$ an $0.25$, respectively. It can be seen that the introduction of surfactant  overcomes the tendency of oil and water to phase separate, resulting in a final equilibrium state which consists of droplets of oil in water. It can also be seen that in the final state the surfactant particles reside predominantly at the oil-water interfaces. Next we performed a simulation in which oil is totally removed from the system. As figure~\ref{fig:two} shows, in the absence of oil-water interactions, the tendency of amphiphile to cause interfaces to self-assemble produces a lamellar structure with long range order to form from an initially homogeneous mixture.

\subsection{Lattice Gas simulations}

The following set of canonical parameters as defined in equation~\ref{eq:hamil} are used consistently throughout our LGA simulations: $\alpha=1.0, \epsilon=2.0, \mu=0.75, \zeta=0.5$. Lattice gas simulations at an oil-water-surfactant composition $0.25$, $0.50$ and $0.25$, identical to that used in the lattice-Boltzmann simulations described above, do not lead to the same final state. Whereas the lattice-Boltzmann morphology shown in Figure~\ref{fig:one} is that of an ordered array of spherical droplets, the lattice gas simulations lead to a phase separated final state. This difference in behaviour is due to the particulate fluctuations inherent in the lattice gas but wholly absent in lattice-Boltzmann models. An array of spherical droplets, once formed, has no hydrodynamic mechanism of coarsening, and so the domain structure must coarsen by droplet coalescence. This coalescence occurs due to Brownian  motion of the droplets. Lattice-Boltzmann simulations are inherently fluctuationless, and without the {\it ad hoc} insertion of noise sources are well known to lack such Brownian motion effects~\cite{bib:em1,bib:wagneryeo}. 

\begin{figure}[htp]
\centering
\includegraphics[width=0.4\textwidth]{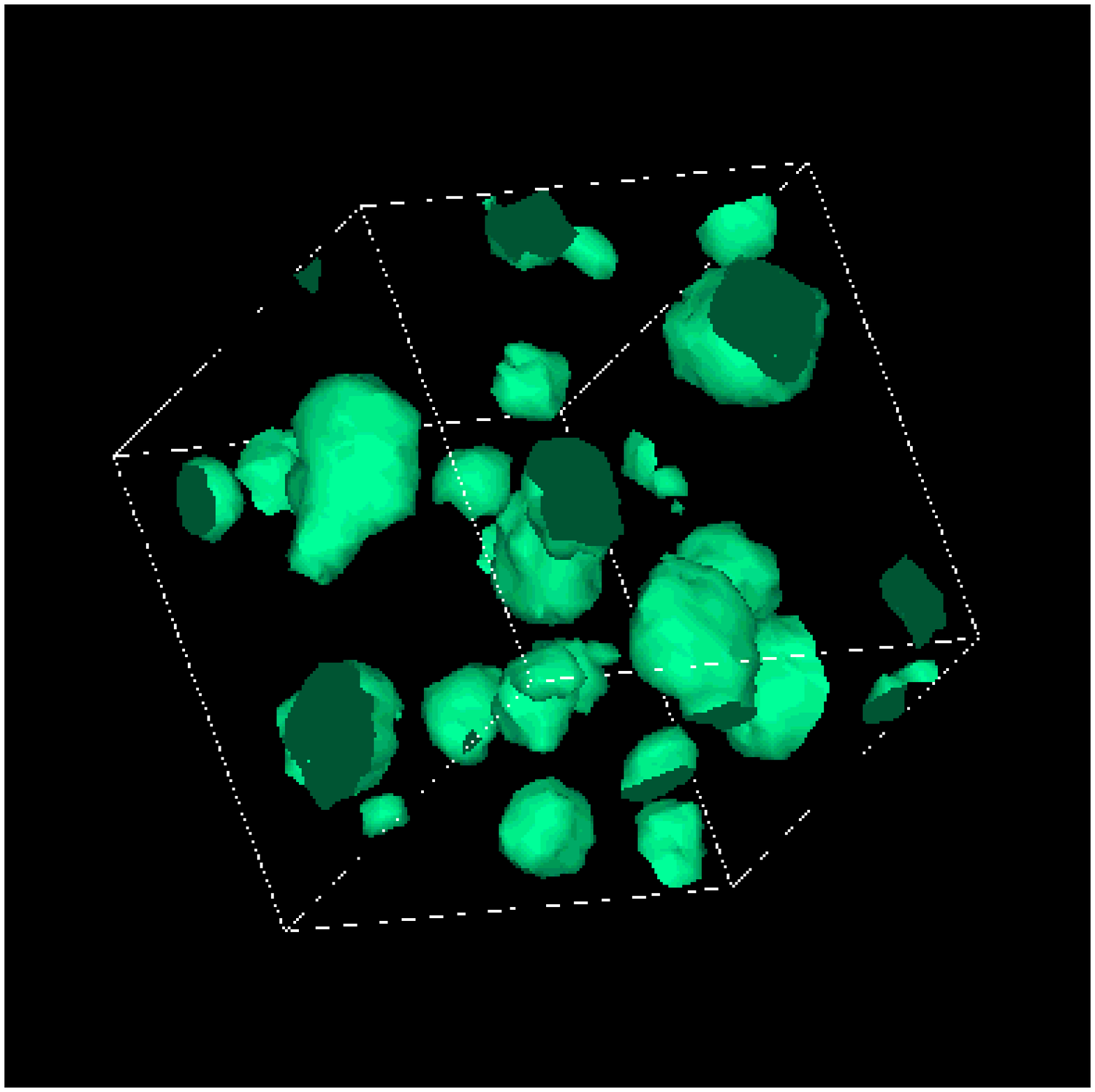}
\includegraphics[width=0.4\textwidth]{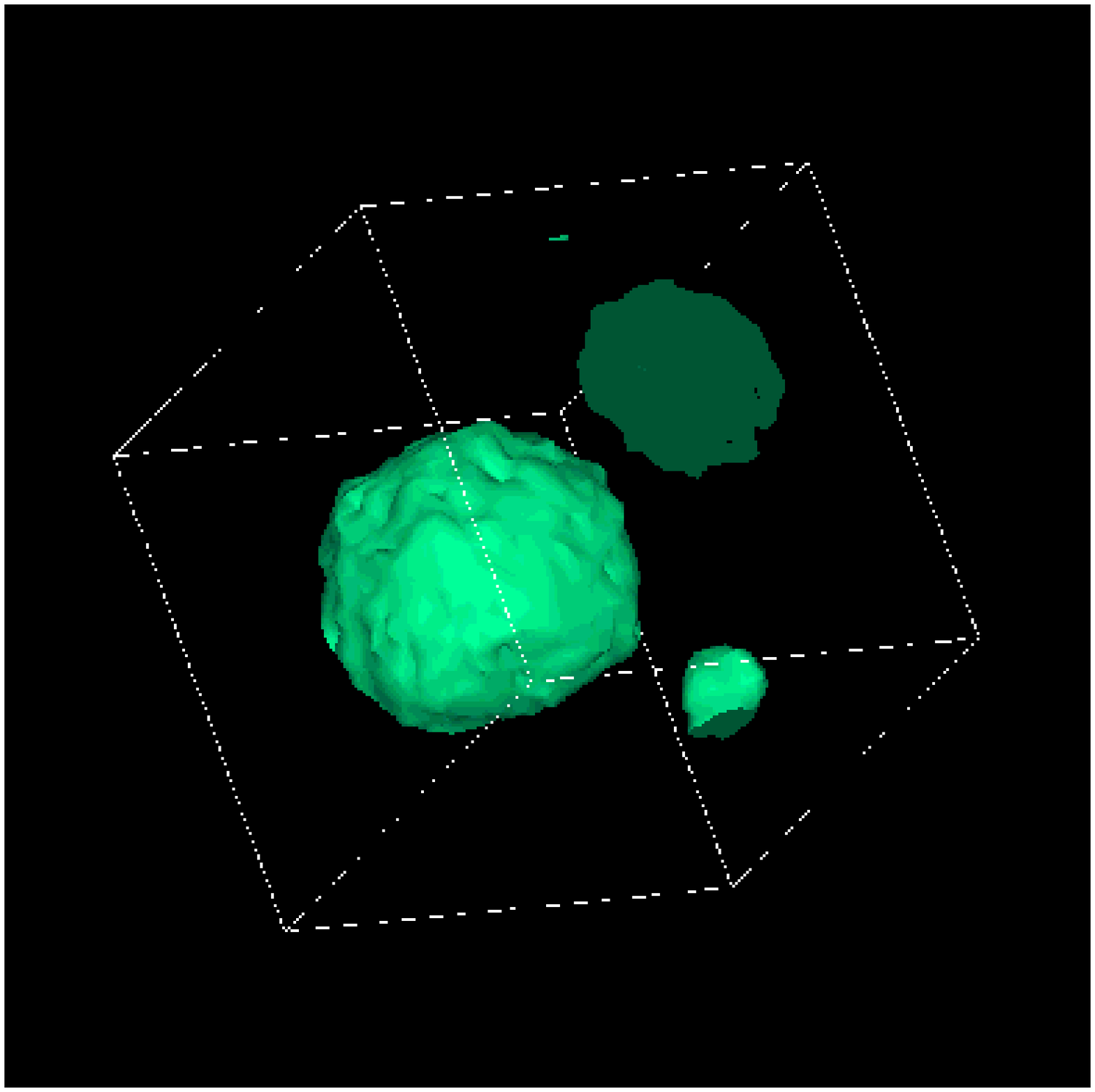}
\includegraphics[width=0.4\textwidth]{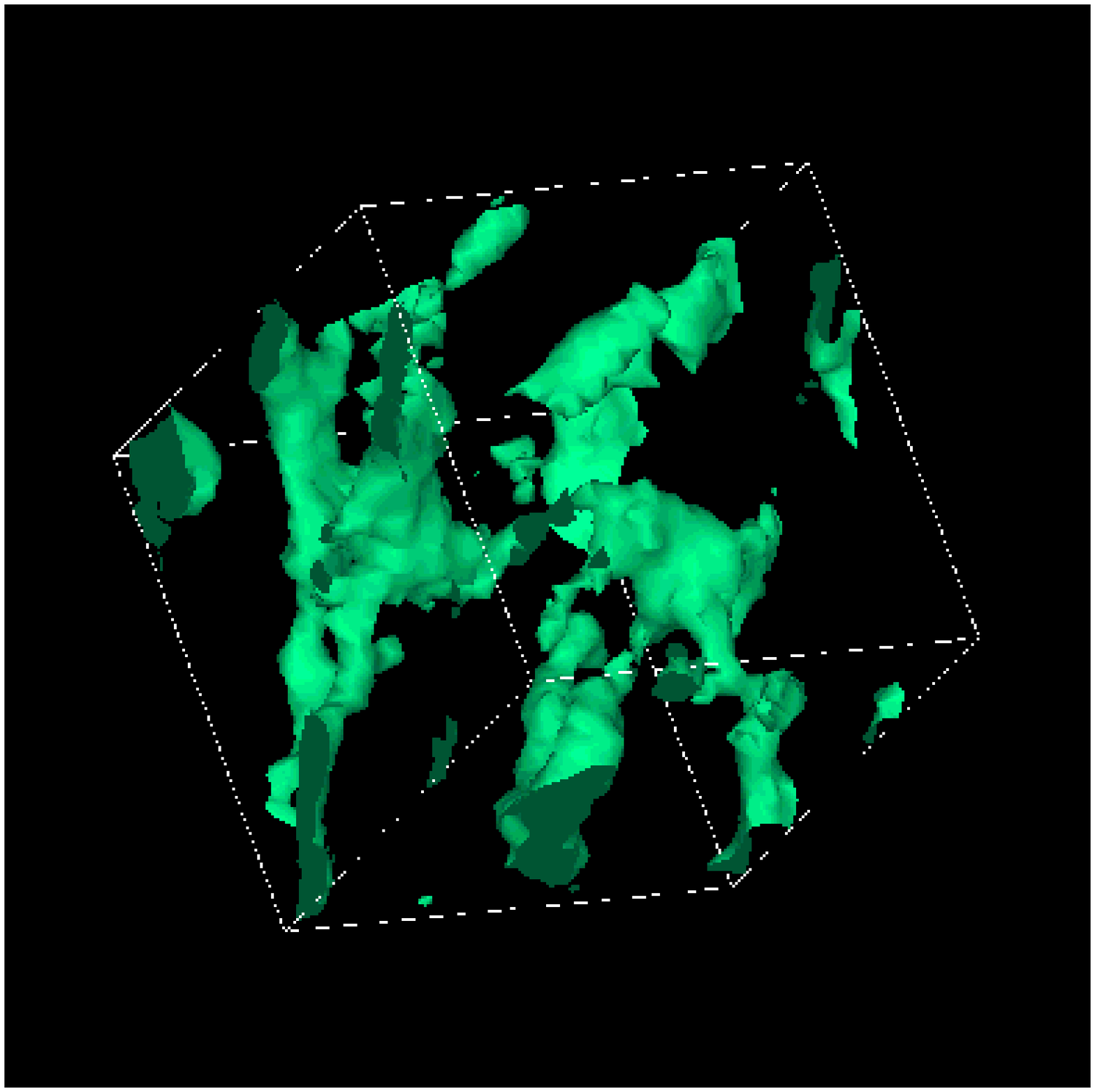}
\includegraphics[width=0.4\textwidth]{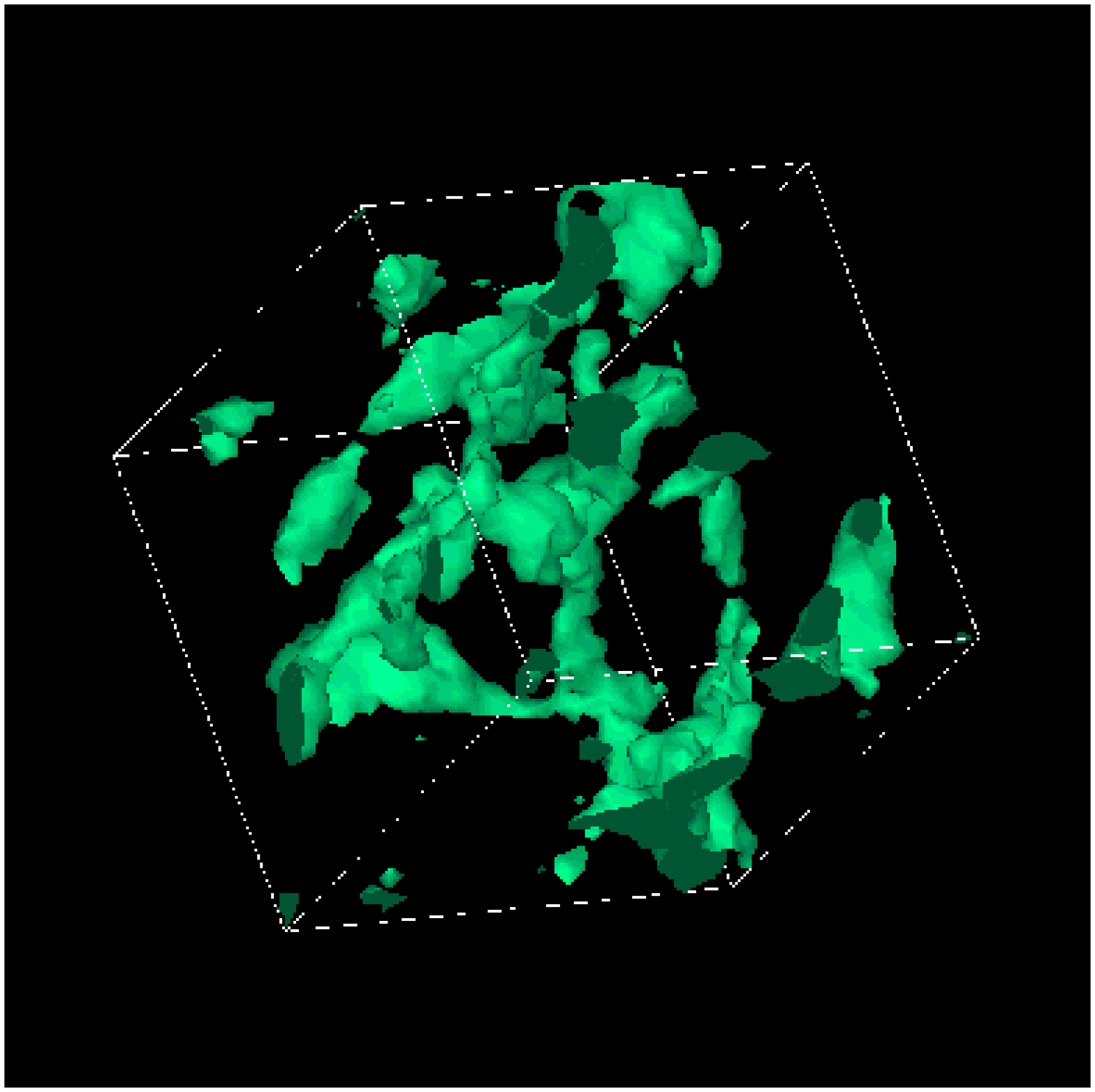}
\caption{\label{fig:three}LGA simulation of binary immiscible and ternary amphiphilic droplet phases. Upper panels show the binary system at time steps $400$ (left) and $6000$ (right). Lower panels show swollen wormlike micellar phase arising from a homogenised initial condition. Isosurfaces show oil concentration at a level of five particles per site. Oil:water:surfactant ratio is $1:19:0$ in the upper panels and $1:10:5$ in the lower panels.}
\end{figure}

In Figure~\ref{fig:three} we show the behaviour of a lattice gas simulation which reproduces a droplet microemulsion phase. The results shown are from two simulations, one in which the average concentrations of water, oil and surfactant are set at $0.95$, $0.05$ an $0.0$, respectively, and one in which the average concentrations of water, oil and surfactant are set at $0.50$, $0.05$ an $0.25$, respectively. It can be seen that the introduction of surfactant overcomes the tendency of oil and water to phase separate. However, unlike the lattice-Boltzmann case the final state does not consist of spherical droplets of oil in water. Instead, the oil forms elongated structures, which we refer to as swollen wormlike micelles. The transition from spherical to non-spherical structures in the presence of surfactant is indicative of a balance between the surface tension and stiffness of the interface with surfactant adsorbed to it. 

\begin{figure}[htp]
\centering
\includegraphics[width=0.4\textwidth]{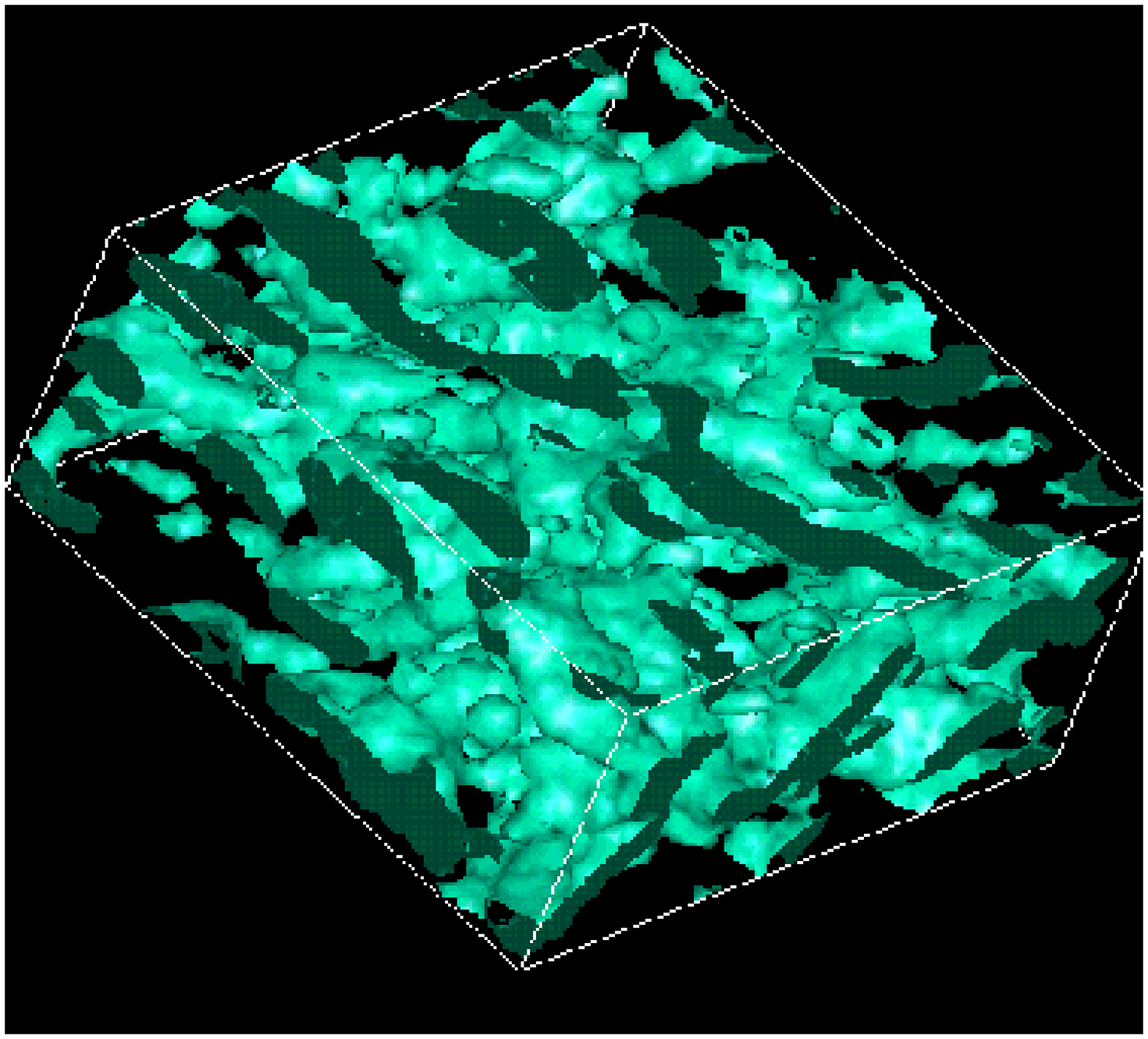}
\includegraphics[width=0.4\textwidth]{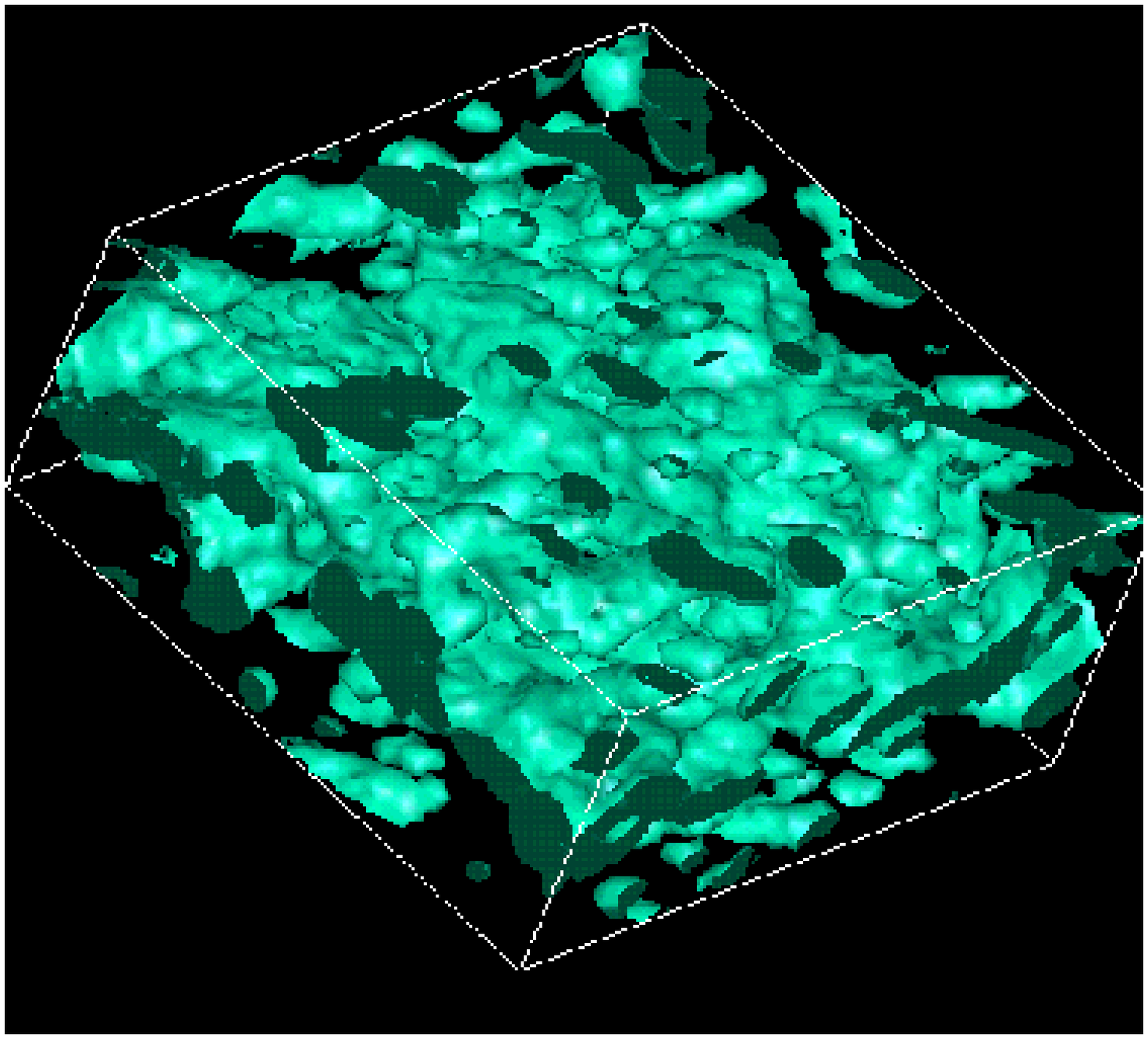}
\caption{\label{fig:four} LGA simulation of binary amphiphilic phases at time steps $400$ (left) and $6000$ (right). Isosurfaces show surfactant concentration at a level of five particles per site. A $64 \times 64 \times 33$ slice through a $64^3$ system is displayed. Oil:Water:Surfactant ratio is $0:3:1$}
\end{figure}

Next we performed a simulation in which oil is totally removed from the system, and the average concentrations of water, oil and surfactant are set at $0.75$, $0.05$ an $0.25$ respectively. Figure~\ref{fig:four} shows that in the early stages of the simulation the surfactant forms a network of wormlike micelles. Later in the simulation the formation of a bilayer (albeit one containing several defects, or holes) is apparent. Thus the final equilibrium state for this system in the lattice-gas is much more dynamic, with surfactant particles forming and reforming a range of aggregated structures. 

\section{Parallelization method and performance}\label{parallel}
The CPU time and memory requirements of ME3D and LB3D scale as $M\times N^D$ where $M$ is the number of discrete velocity vectors, $N$ is the linear size of the system and $D$ is spatial dimension. In two dimensions it is possible to study adequate system sizes and time scales using standard workstations. In three dimensions, however, the serial algorithm quickly becomes prohibitive in terms of computer memory and CPU time even for moderately sized systems. Fortunately, an important feature of the  methods is their intrinsically parallel structure: using a unified strategy we have implemented  parallel version of both algorithms which have allowed us to perform  large-scale $3D$ studies on massively parallel platforms. In this section we describe our parallelization method and the results of benchmark studies of the two parallel codes.

\subsection{Parallelization strategy}
\begin{figure}[htp]
\centering
\includegraphics[width=0.8\textwidth]{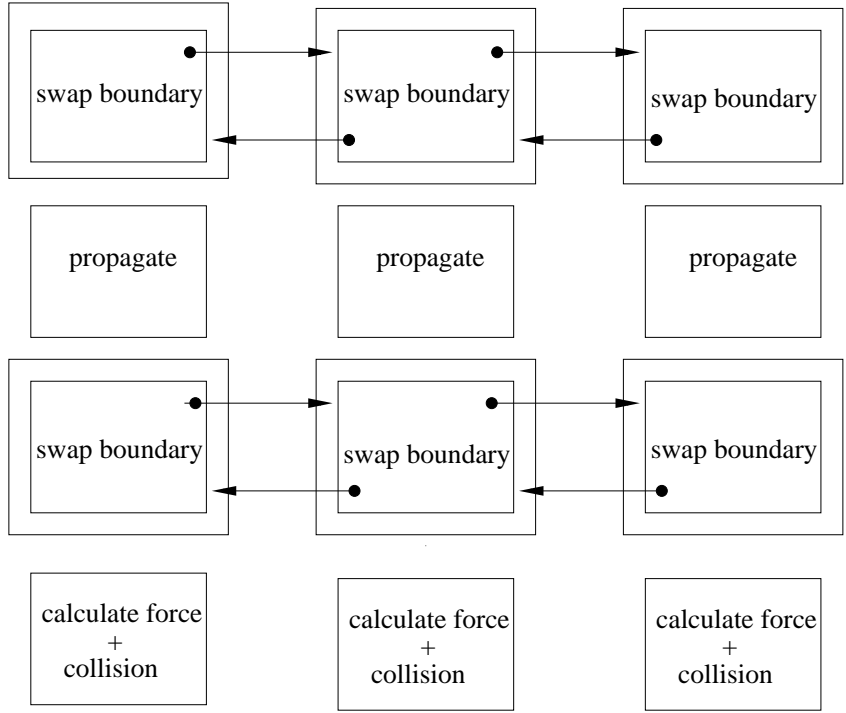}
\caption{\label{fig:flowchart} Flow chart of the parallel LGA and LBE algorithms (see text for further discussion).}
\end{figure}

Parallelization was performed utilizing the single program multiple data (SPMD) model, wherein the same program is executed on all processors, and by means of a domain decomposition strategy. This strategy is very suitable for grid-based and semi-local algorithms like lattice-gas, lattice-Boltzmann, finite-difference and finite-element methods. The underlying 3D lattice is partitioned into spatial sub-domains and each sub-domain is assigned to one processor. Such a decomposition may be done in one-dimension (slices), two-dimensions (rods) or three-dimensions (boxes). We have used  box decomposition which offers maximum flexibility and also results in a minimum surface:volume ratio of sub-domains, hence minimizing  communication overheads. Each processor is responsible for the particles within its sub-domain and performs exactly the same operations on these particles. Two rounds of communication between neighbouring sub-domains are required: at the propagation step, where particles on a border node can move to a lattice point in the sub-domain of a neighbouring processor, and in evaluating the forces, fields and fluxes. By using a ghost layer of lattice points around each sub-domain, the propagation and collision steps can be isolated from the communication step. Before the propagation step is carried out the values at the border grid points are sent to the ghost layers of the neighbouring processor and  after the propagation step an additional round of communication is performed to update the ghost layers. This additional round of communication is  required because of the presence of  non-local interactions in the model whose computation requires (the updated) occupation numbers/distribution functions at neighbouring sites. Figure~\ref{fig:flowchart} shows schematically one step of the parallel algorithm. In the current implementation of the codes we have only considered nearest-neighbour interactions. However, an efficient generalization to long-range interactions of the parallel algorithm, which makes use of parallel fast-Fourier transform (FFT) algorithms, is possible and has currently been implemented in $2D$~\cite{bib:nelido}.

The parallel codes are written in standard FORTRAN90, and make use of a number of features of that language that are object-oriented in spirit. Essentially all of the subprograms used in the codes are {\it encapsulated}, in that they accept arrays of any size specified at execute time.  Many of them are also {\it overloaded} in that they can accept arguments of a variety of different ranks and types, and do the right thing in each case. The codes also make use of {\it user-defined types} to represent the data. The codes utilise the standard message passing interface MPI for synchronization and communication between processors~\cite{bib:MPI}. Most of  the MPI calls, however, have been embedded in {\it wrapper} functions to allow relative ease of porting the code to any other message passing protocol if desired.  To port to another parallel language, the MPI calls can simply be replaced by their equivalents in the alternative protocol.

\subsection{Parallel performance}

The parallel performance of the codes was benchmarked on two parallel platforms with totally different architectures: the CRAY T3E 1200E and the SGI O2000. The Cray T3E system is a distributed memory Massively Parallel Processor (MPP) machines (Dec Alpha EV6 processors running at 600 MHz, with $256$ Mbytes of memory each), whereas the SGI O2000 machines are Shared Memory Parallel (SMP) machines with cache-coherent Non-Uniform Memory Architecture (ccNUMA) (192 MIPS R10000 processors running at 195 MHz, with 768 MB of high-speed cache memory). All benchmark simulations were performed on $64^3$ systems using the PFCHC lattice with coordination number $24$ for $500$ time steps. The ME3D lattice vectors are the $24$ nearest-neighbours of the site (0,0,0,0,0), projected to $3D$ plus two rest particles. In LB3D one rest distribution is used.

Fluid densities at each lattice site and surfactant dipole moments were written out periodically at every $50$ time steps of the simulations. The presence of surfactant interactions and dynamics significantly increases the  memory requirements and execution time of both algorithms. Since in ME3D the occupation numbers are boolean operators the absence of surfactant at a given site can be easily detected and evaluation of the surfactant-dependent portions of forces and fluxes at such sites omitted. This feature is built into the ME3D code such that the overall computation time of the algorithm depends on the average concentration of surfactant in the fluid, and is greatly reduced when little or no surfactant is present in the system. Since the LB3D ``occupation numbers'' are real variables, implementing a similar feature in the LB3D algorithm is less straightforward and was not attempted. Instead, we implemented two parallel versions of the LB3D algorithm, one for binary immiscible systems and one for ternary amphiphilic systems. Parallel performance of both binary and amphiphilic version of the two algorithms  was benchmarked on the distributed memory CRAY T3E 1200E massively parallel platform. To make a fair comparison possible between the performance of the amphiphilic LB3D and ME3D codes we used the same concentrations of surfactant in benchmarking these codes.

In Tables~\ref{tab:baseline1} and~\ref{tab:baseline2} we compare the performance (number of time steps per minute) of the binary and the ternary codes, respectively. It can be seen that on the Cray T3E the LB3D codes are $\sim 5$ times faster than the  corresponding ME3D codes, regardless of the number of processors used. The only situation in which the ME3D code is able to seriously compete with the LB3D code is for the binary system on the O2000 architecture, where LB3D is only $10\%$ faster. The LB3D code is approximately twice as fast for the ternary system on this architecture. This is consistent with our observation above that the ME3D code must compute a large number of components of tensor fluxes and fields in the ternary case. It is likely that the better cache handling properties of the O2000 enhance the performance of the computation of these quantities, and that in the binary case where these calculations are unnecessary the ME3D code is able to maximise its competitiveness with LB3D. 

We note that since ME3D is an inherently stochastic algorithm the calculation of physical quantities such as surface tension, diffusion coefficients, etc. with this scheme involves additional ensemble-averaging over a set of initial conditions in order to eliminate statistical noise, which is absent in the lattice-Boltzmann method. This increases the effective computational time of ME3D by, roughly, a factor $4$ or $5$.

\begin{table}
\caption{Performance (time step/minute) of the ME3D and LB3D codes on 
         Origin 2000 Silicon Graphics\label{tab:baseline1} } 
\begin{tabular}{llllll}
\hline
\\
time steps/minute & $8$ CPU & $16$ CPU & $32$ CPU& $64$ CPU \\
\hline
LB3D (binary)  & $19.0$ & $32.6$ & $54.2$ & $110.4$   \\
LB3D (ternary) & $7.7$ & $16.7$ & $34.30$  & $59.8$  \\
\hline
ME3D (binary)  & $7.1$ & $20.1$ & $51.76$ & $100.0$   \\
ME3D (ternary) & $5.3$ & $8.0$ & $19.45$  & $30.36$  \\
\hline
\end{tabular}
\end{table}

\begin{table}
\caption{Performance (time step/minute) of the ME3D and LB3D codes on 
         CRAY T3E \label{tab:baseline2}} 
\begin{tabular}{llllll}
\hline
\\
time steps/minute & $8$ CPU & $16$ CPU & $32$ CPU
& $64$ CPU \\
\hline
LB3D (binary)  & $23.0$ & $44.3$ & $80.5$ & $133.3$   \\
LB3D (ternary) & $9.8$ & $19.3$ & $36.0$  & $58.1$  \\
\hline
ME3D (binary)  & $1.66$ & $4.00$ & $5.50$ & $28.57$   \\
ME3D (ternary) & $0.7$ & $2.7$ & $6.1$  & $11.4$ \\
\hline
\end{tabular}
\end{table}

Many analyses of the parallel performance of codes rest on plots of the so-called speedup of the code. We have measured a sequence of speeds $S_{N_i}$, measured in time steps per minute for a sequence of parallelisations across$N_i$ processors. The speedup is usually defined as:
\begin{equation}
SU_i = \frac{S_{N_i}}{S_{N_0}},
\end{equation}
where $N_0$ is the smallest number of processors across which the system size chosen will run. Plots of speedup are therefore critically dependent on the choice of $N_0$. For $N_0$ too small the codes' performance will be completely dominated by memory bandwidth effects, where the bottleneck is moving data in and out of on-processor memory, irrespective of processor speed and inter-processor bandwidth and latency. As one increases the number of processors the amount of data being moved through the processors' cache is reduced until some critical unit of data will fit entirely into the on-processor cache. At this point the bottleneck becomes either computational speed, or communication time. In order to separate the parallel performance of our codes from dependence on the unrepresentative behaviour of simulations on small numbers of processors we plot the parallel efficiency:
\begin{equation}
E_i = \frac{S_{N_{i+1}}}{S_{N_i}} \frac{N_{i}}{N_{i+1}}.
\end{equation}
This quantity is equal to one if doubling the number of processors doubles the performance (linear scaling), greater than one if the performance is more than doubled (superlinear scaling), and less than one if the performance is less than doubled (sublinear scaling). 

\begin{figure}[htp]
\centering
\subfigure[]{\resizebox{0.4\textwidth}{!}{\includegraphics{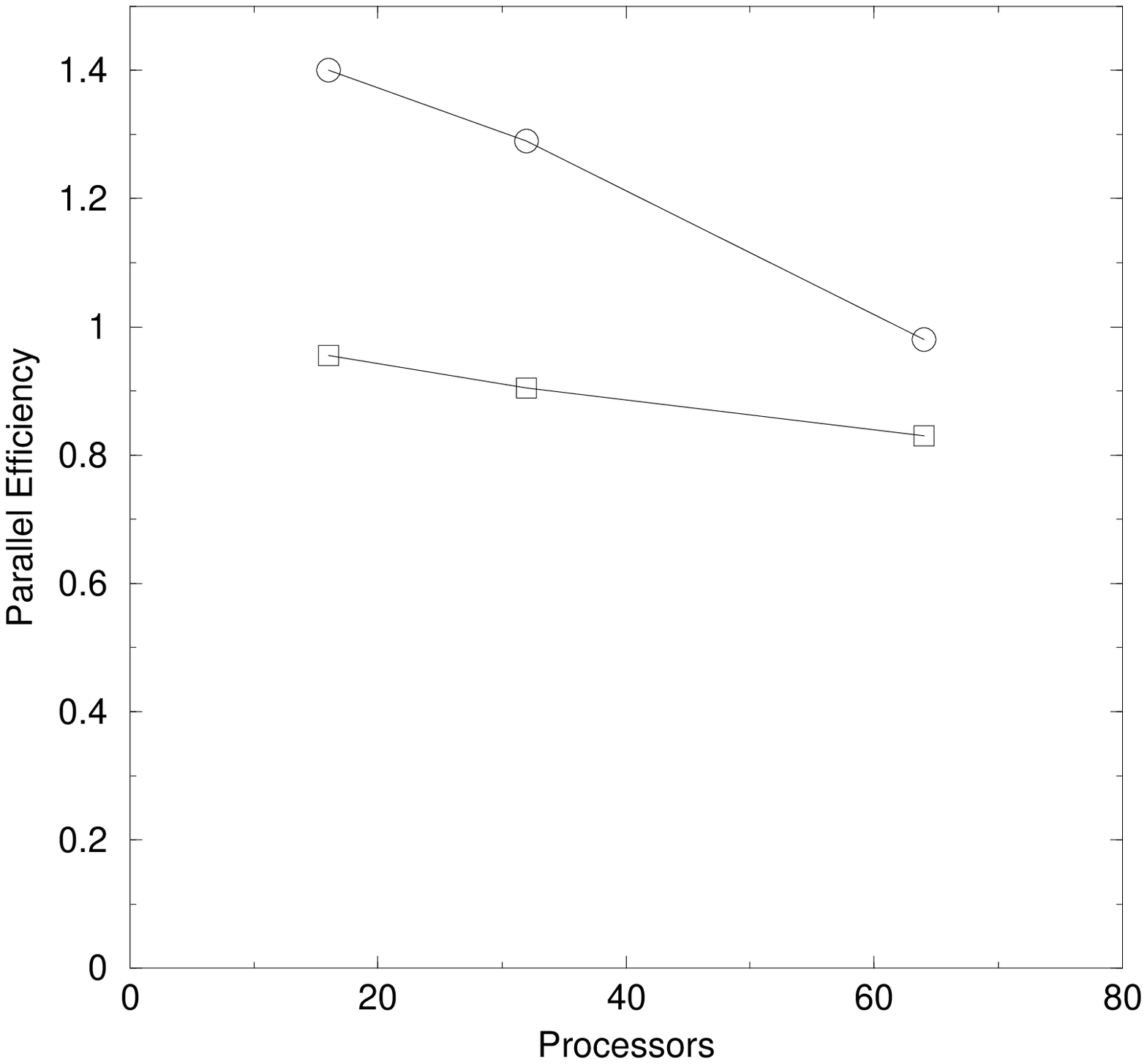}}}
\subfigure[]{\resizebox{0.4\textwidth}{!}{\includegraphics{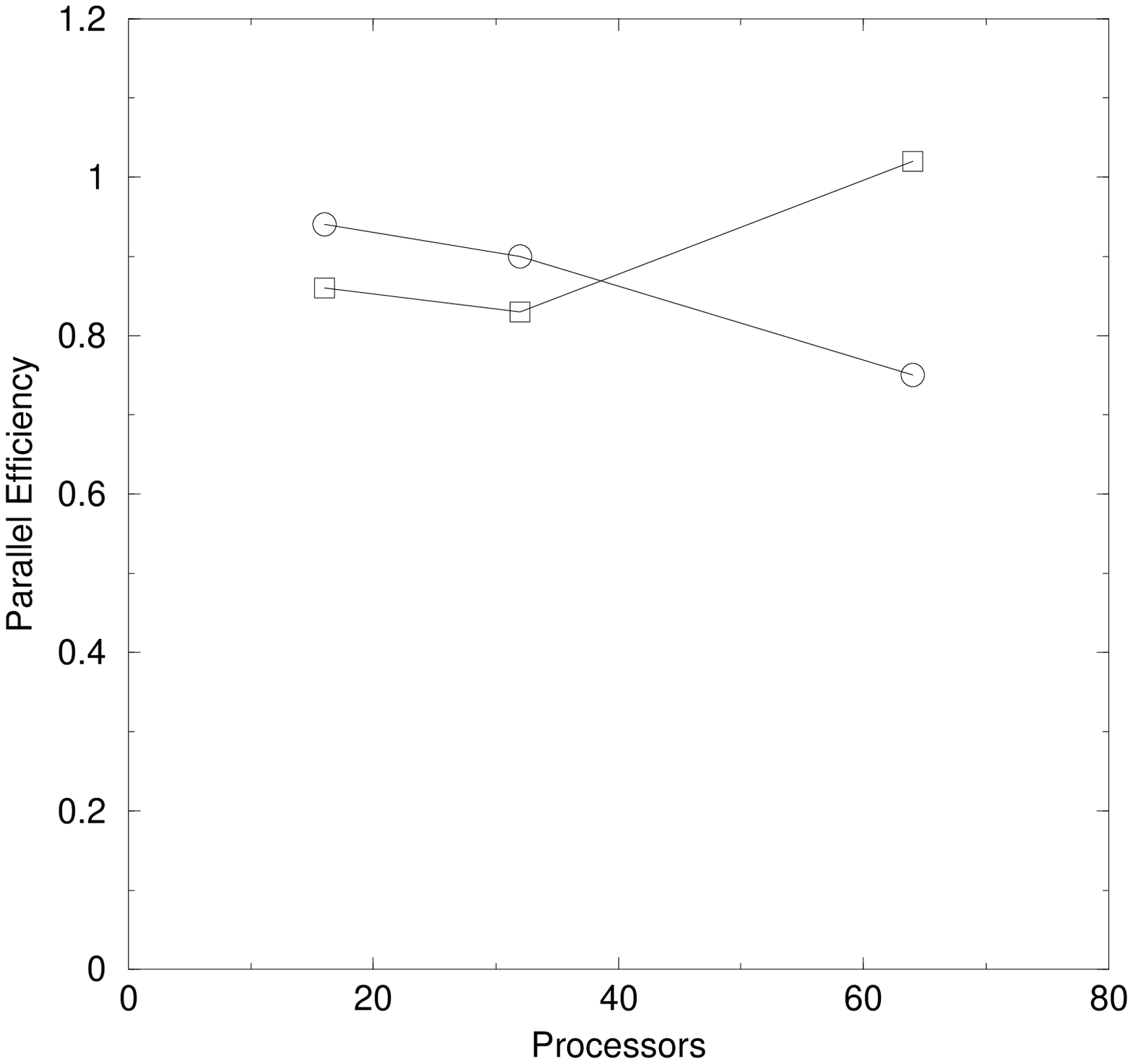}}}
\caption{\label{fig:bineff}Parallel efficiency (vertical axis) vs. number of processors (horizontal axis) for binary immiscible fluid simulations using the LB3D and ME3D codes measured on (a) Origin 2000 and b) CRAY T3E 1200E. Circles and squares show ME3D and LB3D performance respectively}
\end{figure}

The parallel efficiency of the ME3D and LB3D codes for the binary system is shown in Figure~\ref{fig:bineff}. The superlinear scaling of the ME3D code shows that it is significantly influenced by cache effects on the O2K, as previously supposed from its greater competitiveness with LB3D on this platform and system. The LB3D code has very high efficiency and sublinear scaling on the O2K, indicating a computation dominated code uninfluenced by cache effects. The behaviour on the T3E is more complicated. ME3D shows a high efficiency with sublinear scaling, while LB3D has a critical number of processors above which the efficiency improves. 

\begin{figure}[htp]
\centering
\subfigure[]{\resizebox{0.4\textwidth}{!}{\includegraphics{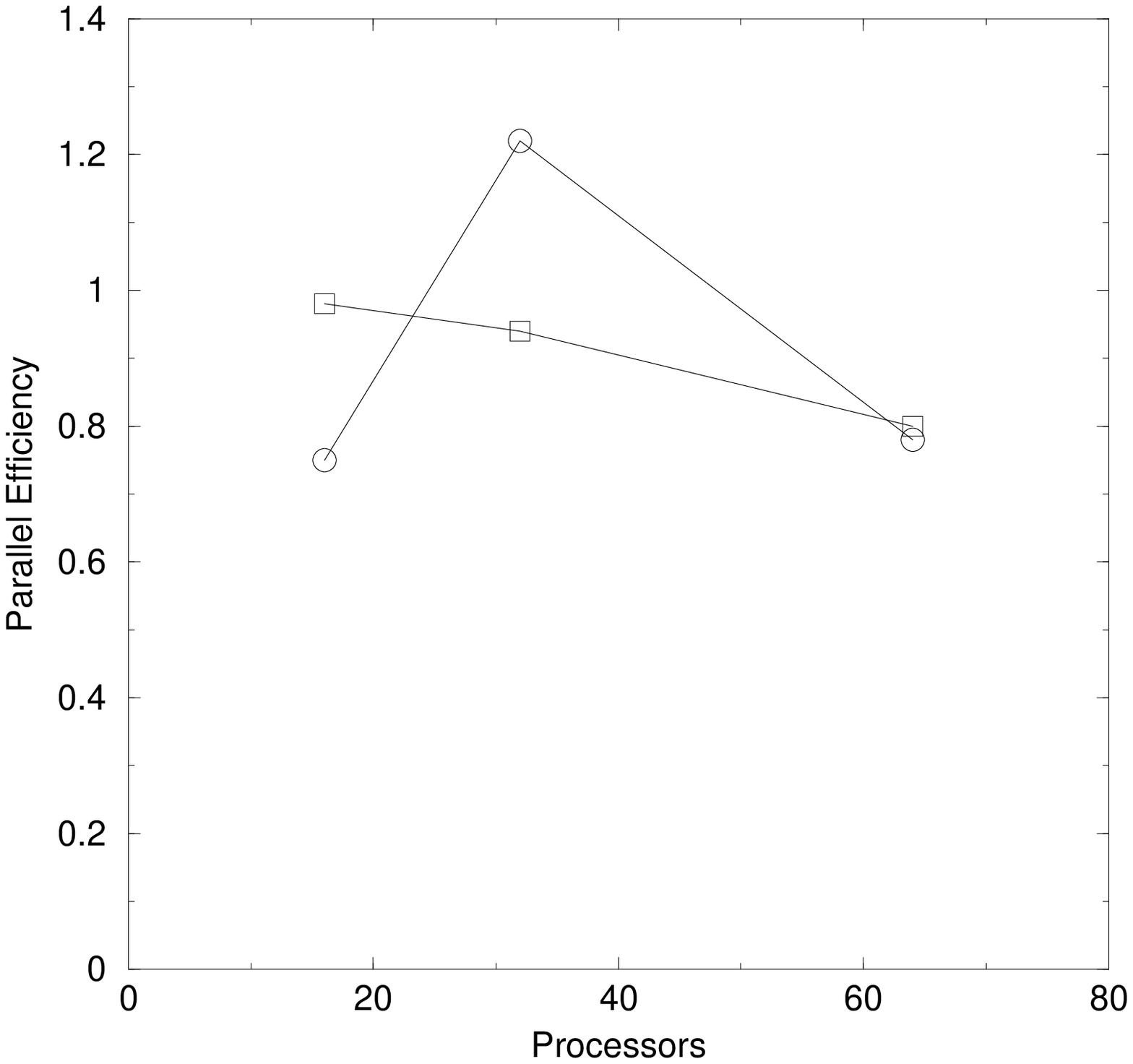}}}
\subfigure[]{\resizebox{0.4\textwidth}{!}{\includegraphics{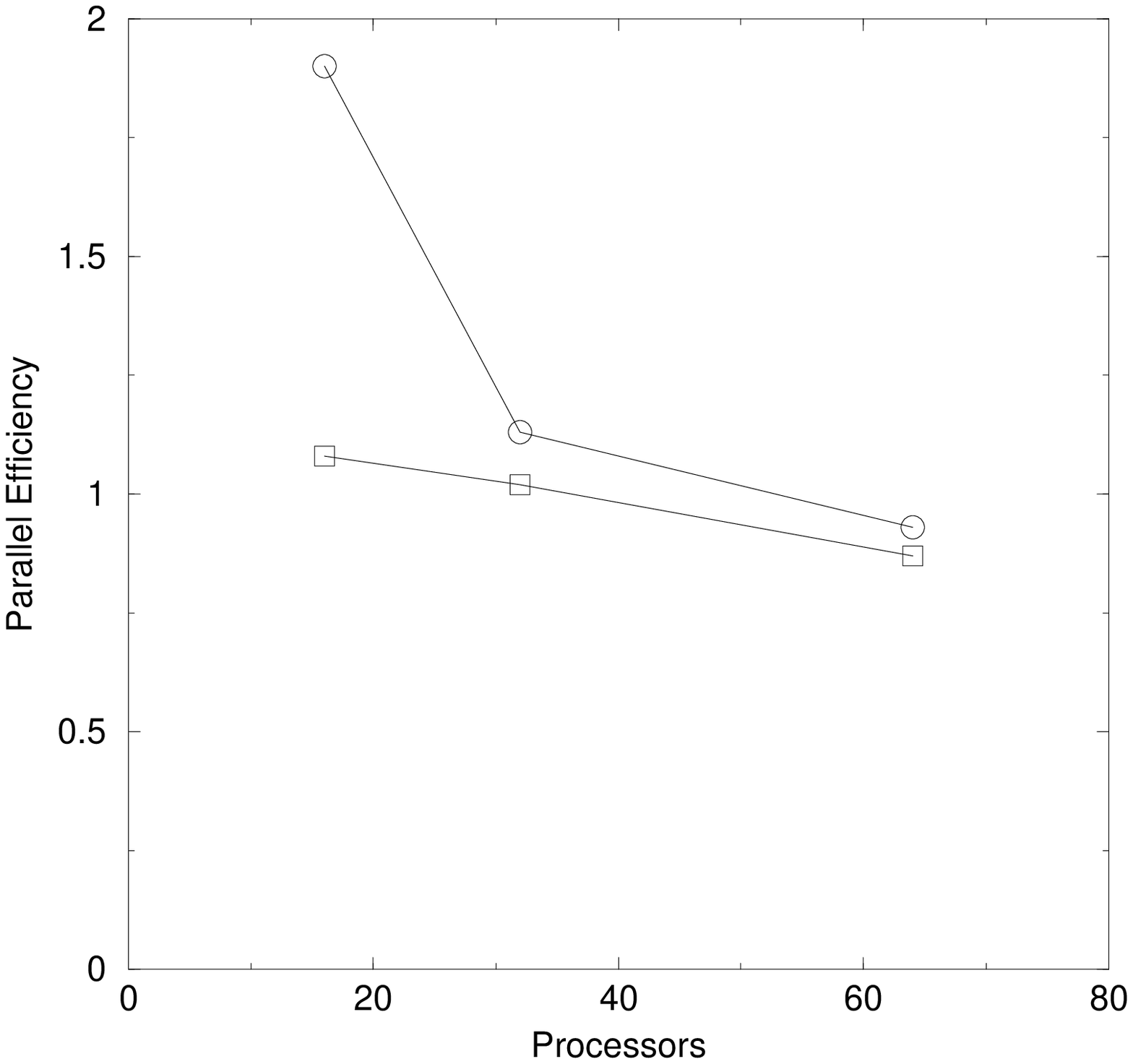}}}
\caption{Parallel efficiency (vertical axis) vs. number of processors (horizontal axis) for ternary amphiphilic fluid simulations using the ME3D and LB3D codes measured on (a) Origin 2000 and (b) CRAY T3E 1200E. Circles and squares show ME3D and LB3D performance respectively.}
\label{fig:terneff}
\end{figure}

For the ternary system LB3D shows high efficiency and sublinear scaling on both platforms, while ME3D shows a peak of superlinear performance at $32$ processors on the O2K, after which the efficiency scales sublinearly, and at $16$ processors on the T3E, beyond which the efficiency stays superlinear (Figure~\ref{fig:terneff}).  This shows that the design of the O2K memory system, with a large 8MB secondary cache, yields significant advantages above 32 CPUs. The T3E
only has a 96 kB secondary cache, but also has several other features,
such as stream buffers and ``E-registers'' to improve memory bandwidth
and reduce latency.  The T3E yields a gradually decreasing advantage with
decreasing data per CPU, the codes' performance still showing possible
memory bandwidth effects with 64 CPUs. It is possible that the T3E's
memory system would require specific optimizations to take advantage of
its more complicated design.

\section{High Performance Visualisation and Computational Steering}\label{steering}

Technological advances are making parallel computers ever more powerful, and the range of problems that they can tackle is continually expanding. However, a `mainframe mentality' persists in the manner of their use which dates from the nineteen-sixties. They are typically operated as batch-processors sealed off from their users in some way - usually system time must be reserved days in advance\cite{bib:Measuresetal}. The purpose of this method of operation is to try to maximise the flow of work through a heavily-loaded shared resource. However, this approach overlooks the fact that many jobs ultimately produce negative results. 

A common feature of both our LGA and LBE models is the presence of a set of parameters which can be adjusted to produce the desired amphiphilic fluid phenomenology, and/or numerical stability properties. A prerequisite of the use of these methods is the performance of large parameter space searches, which typically involves massively parallel ``taskfarms'', in which each processor independently runs a small system with a different set of simulation parameters. Such searches again require the use of massively parallel centralised computing resources. 
 
Were there the capability to visualise the state of running programs in essentially real time many of the aforementioned failures might be detected at an early stage, freeing up the resources for another purpose. In some cases failure might be averted simply by modifying model parameters and feeding them back into the application in progress. Allowing scientists the freedom to exploit their intuition interactively can greatly reduce the computation time required to get results. {\it Computational steering} is an efficient method for conducting such searches which allows the user to home in to the desired phenomenology by ``steering''  his or her way in the parameter space of the models.

Much research effort has been applied to developing computational steering environments for parallel programming (e.g. \cite{bib:beazlom, bib:VIPAR1, bib:CUMULVS, bib:cactus}). Typically these involve a visualisation component, a communication component, and some sort of model for parallel computation. There are usually tens of thousands of lines of source code that need to be ported to one's own system. This can be intimidating to a potential user, giving the impression that much work will have to be done. This need not be the case and, indeed, the approach we describe here is bare-bones. We do not provide any parallel programming model, but we support the industry-standard MPI library\cite{bib:MPI} (and also OpenMP). We do not specifiy any particular visualisation system, but have found that our code fits easily into existing graphics systems, such as AVS\cite{bib:avs}. All we provide is the `glue' for connecting the parallel program to visualisation code using three simple subroutines. The advantage of this
minimalistic approach is ease of implementation. The effort required to make an existing parallel program steerable is very small, assuming that the operation of the program is well-understood.

\begin{figure}[htp]
\vspace{0.5cm}
\begin{center}
\begin{minipage}{7cm}\leavevmode
\epsfxsize=7cm
\epsfbox{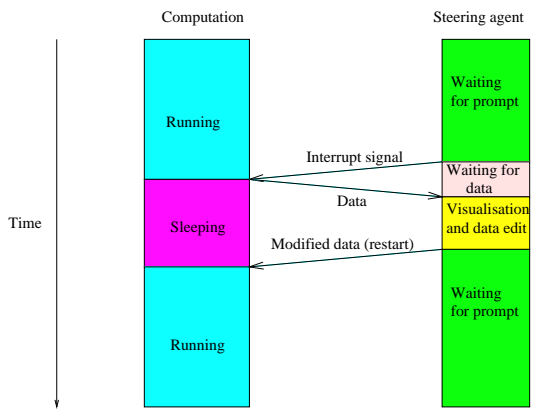}
\end{minipage}
\caption{A basic steering model \label{fig:flowdiag}. See text for more details.}
\end{center}\end{figure}

Our model for computational steering consists of two elements: a parallel application and a steering agent. The steering agent provides a user interface for visualisation and command entry, and may run on a separate machine from the application. When prompted by the user, the agent sends a signal to the application causing it to pause, at a suitable point in its execution, and then send back a snapshot of its data. The data is then visualised and explored by the user and in due course a restart signal is sent back to the application, possibly accompanied by some modified parameters (figure~\ref{fig:flowdiag}). 

The commercial visualisation package AVS (both AVS5 and AVS/Express) was used for all the figures in this paper and for much of our work on steering. As one proceeds to large ($128^3$ and $256^3$) datasets, generating and rendering complex isosurfaces becomes a computationally demanding task. In particular, one requires that the visualisation can be generated rapidly enough that one can interactively rotate, zoom, and otherwise interrogate it. What this means in practice is that for a given visualisation system and type of data (which determines the complexity of the isosurface) there is a maximum data size above which one cannot usefully perform visualisation.

We found that, for generating and viewing isosurfaces from typical amphiphilic fluid datasets larger than around $64 \times 64 \times 64$, the open-source package VTK (www.kitware.com) was much more responsive than AVS. VTK's provision of interfaces to several languages made scripting and rapid prototyping of the visualization networks only marginally more onerous than using AVS's very user friendly drag-and-drop graphical programming interface.

Various parallel solutions to bottlenecks in the visualisation pipeline have been proposed and implemented. In the context of AVS, the rendering bottleneck for large datasets was addressed by the AVS/Express MPE, which implements parallel rendering on multiple graphics pipes. However in our testing of AVS/Express MPE on an SGI Onyx-2 system with 2 pipes we observed no rendering speedup whatsoever. The isosurface (and, in principle, other graphics objects) generation bottleneck was addressed for AVS (and, again in principle, for other visualisation systems) by the VIPAR project~\cite{bib:VIPAR1}. This project implements the parallelisation of graphical object generation routines (``modules'' in the context of AVS). This system is currently undergoing active development by the Manchester Visualisation Centre. Parallel visualisation for VTK is also currently under development~\cite{bib:PARVTK}. 

\subsection{A case study: locating the spinodal point of lattice gas binary immiscible fluids}

As a concrete example of the utilisation of our nascent steering techniques we describe the location of a particular point in the lattice-gas models' phase space using steering techniques. These techniques have also been used to good effect in our lattice-Boltzmann model; however, for reasons of space we confine ourselves to the description of just one example.

\begin{figure}[htp]
\centering
\begin{minipage}{7.1cm}
\begin{minipage}{3.5cm}\leavevmode
\epsfxsize=3.5cm
\epsfbox{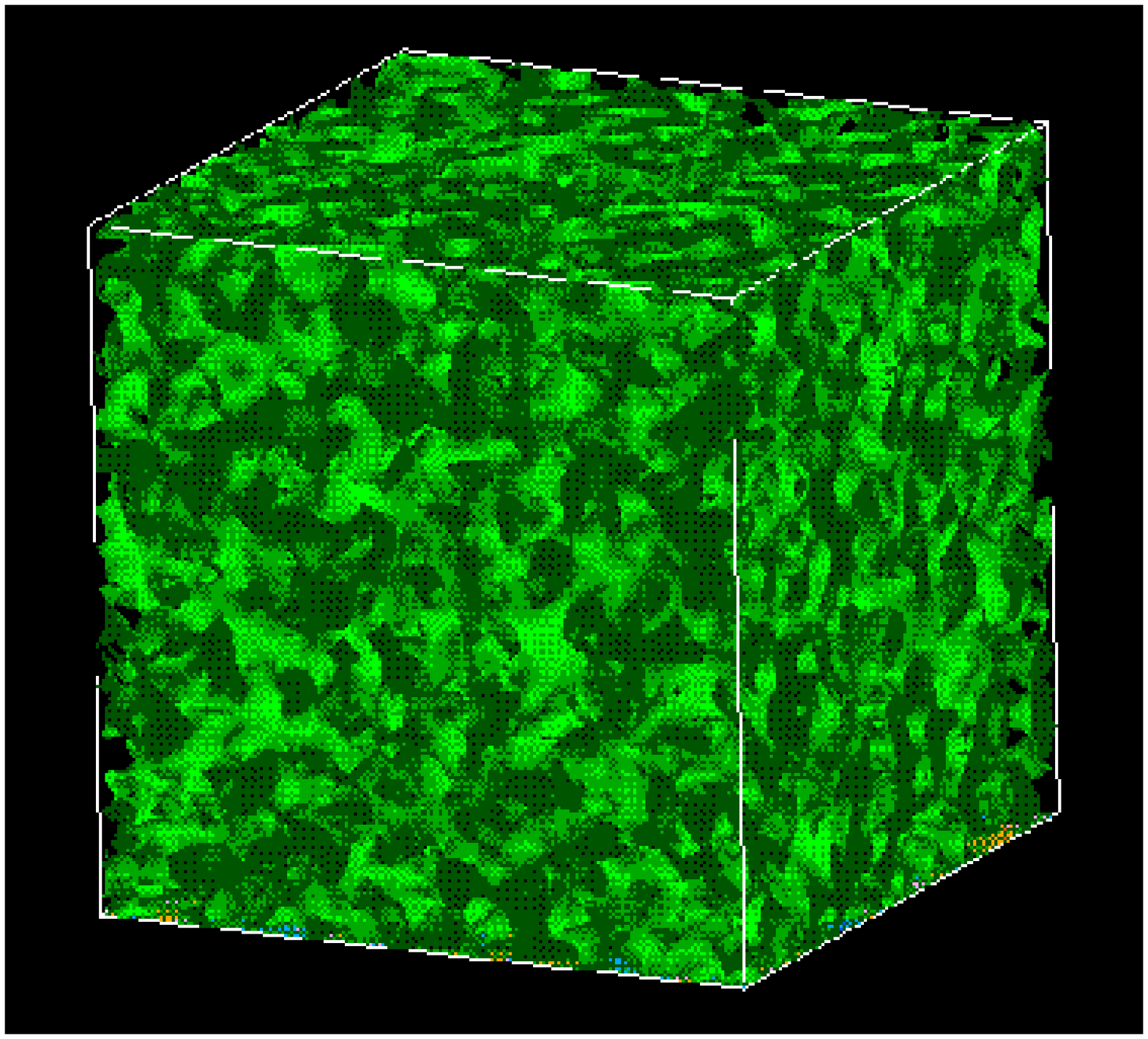}
\end{minipage}
\hfill
\begin{minipage}{3.5cm}\leavevmode
\epsfxsize=3.5cm
\epsfbox{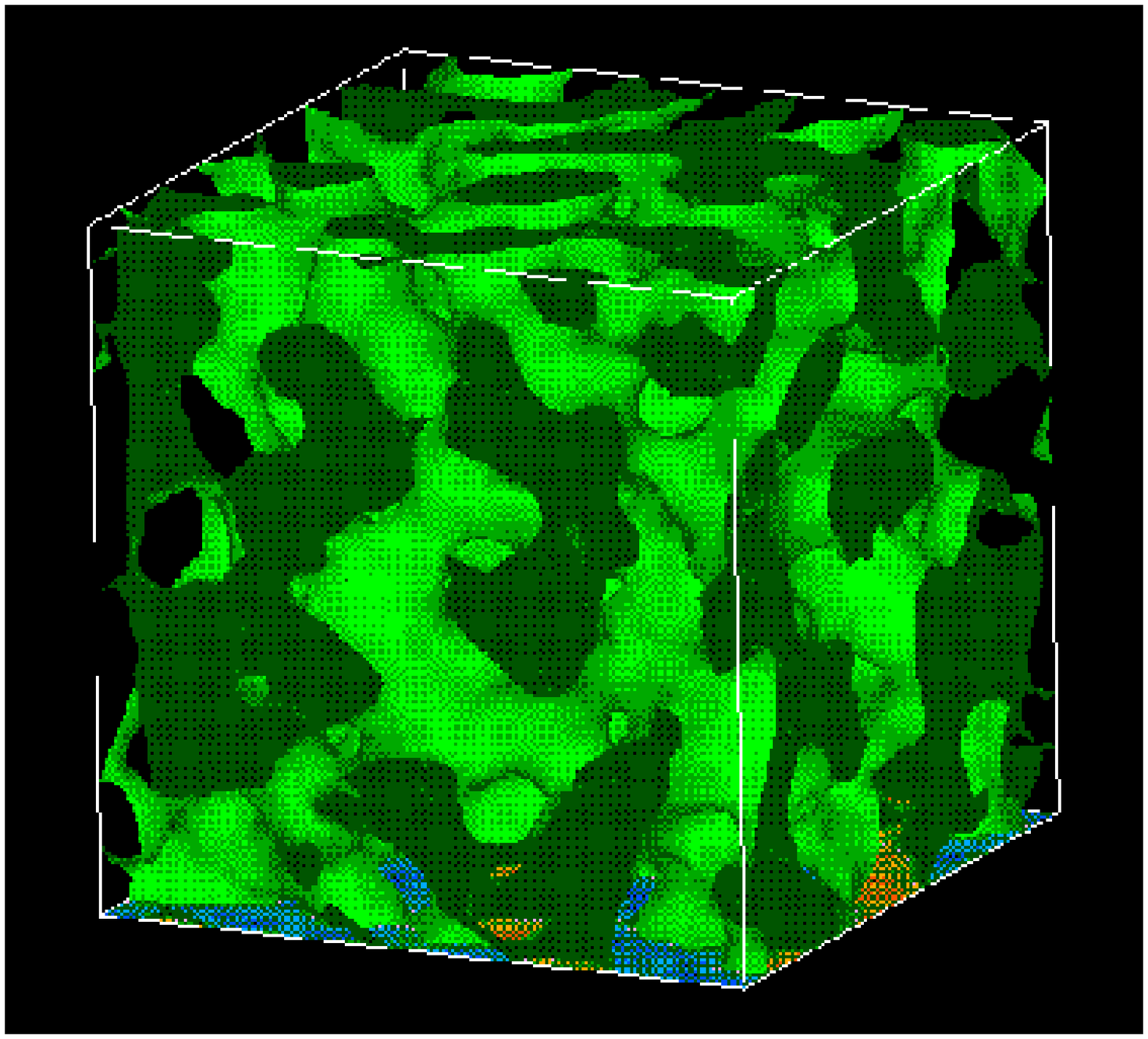}
\end{minipage}
\vspace{0.5cm}
\begin{minipage}{3.5cm}\leavevmode
\epsfxsize=3.5cm
\epsfbox{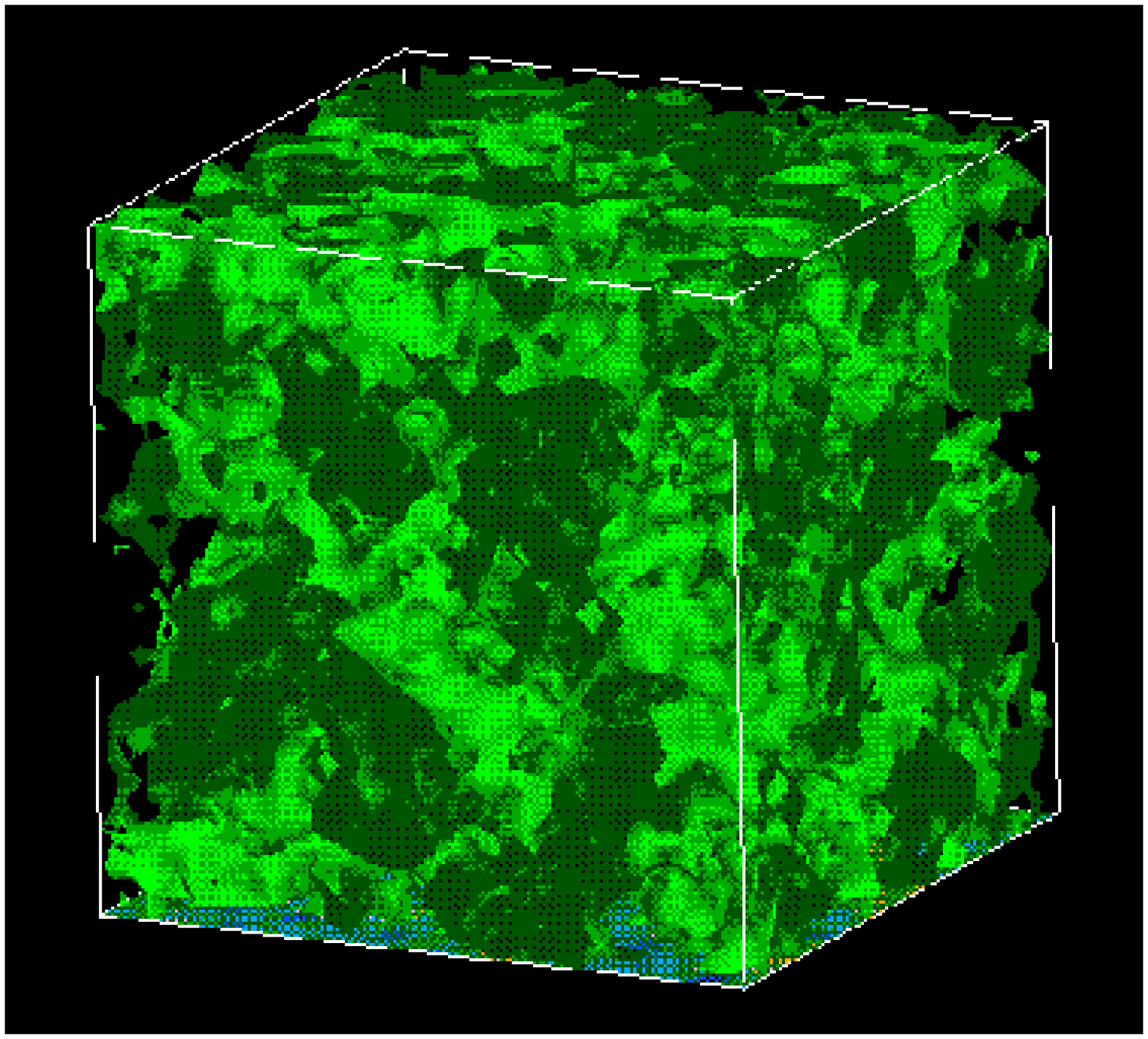}
\end{minipage}
\hfill
\begin{minipage}{3.5cm}\leavevmode
\epsfxsize=3.5cm
\epsfbox{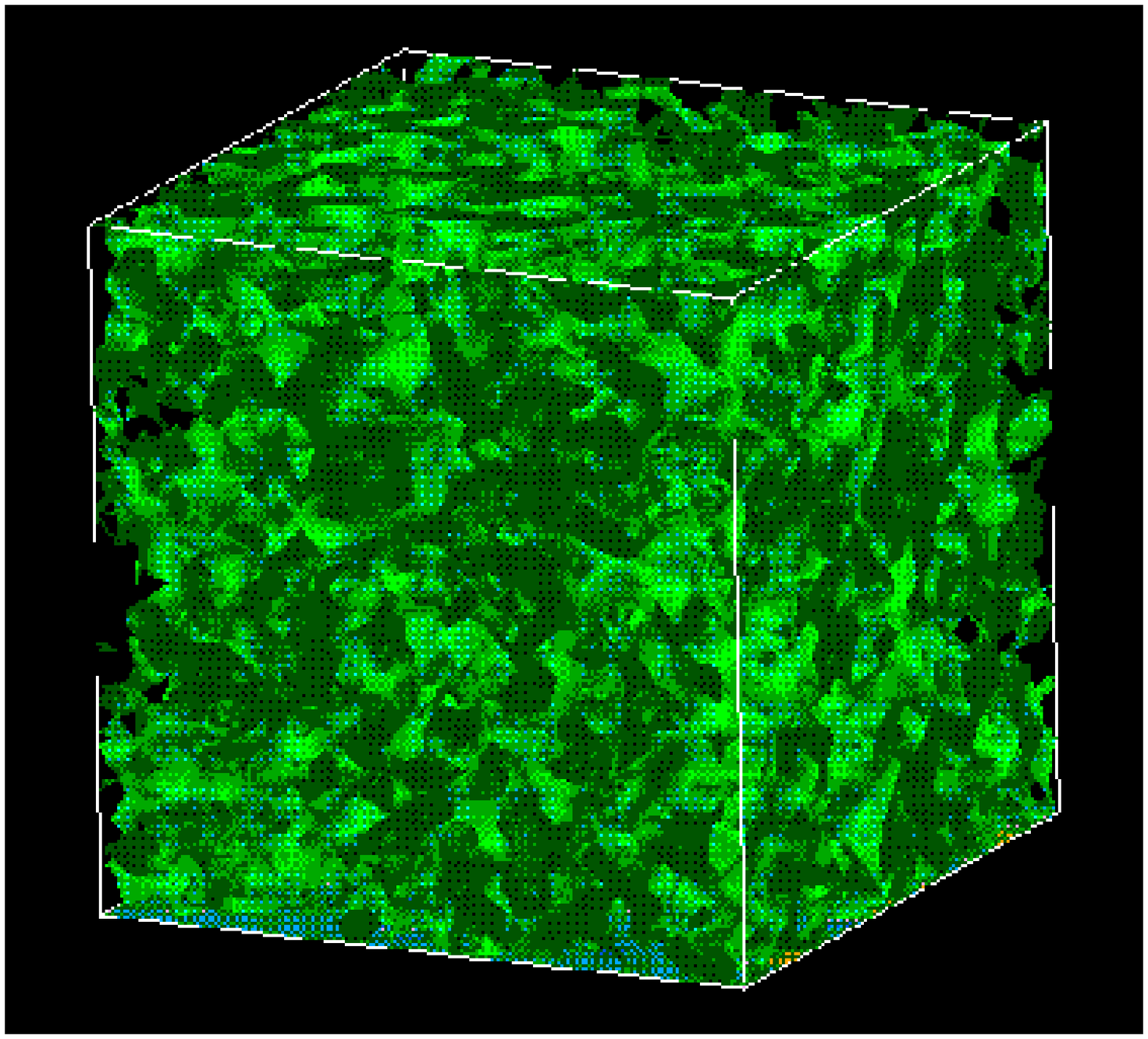}
\end{minipage}
\end{minipage}
\caption[fig:cycle]{The cycle of observations in a
  steered simulation to determine the spinodal point of an immiscible
  fluid. Top left: initial homogenous mixture below spinodal
  point. Top right: Clear phase separation indicating system is below
  spinodal point. Bottom right: System remixing after temperature is
  shifted above spinodal point. Bottom left: System remixed ready for
  the next temperature change.\label{fig:cycle}}
\end{figure}

An initially homogenous mixture of oil and water below a critical temperature referred to as the {\it spinodal point} will spontaneously phase separate. Our lattice-gas model is capable of simulating this behaviour and computational steering was used to locate the spinodal point. Previously such points in the phase diagram of the model have been found using large task-farm parameter searches. Such searches are computationally expensive and time consuming. Additionally, one is limited to a system which will fit on a single processor and so finite size effects may distort the results obtained. In the course of a single simulation lasting under an hour the system's temperature was repeatedly raised above and lowered below the spinodal point and the behaviour observed by direct visualisation until the desired accuracy was obtained. The sequence of events observed in a cycle of simulation is shown in figure~\ref{fig:cycle}. The use of steering meant that the simulation could be stopped when the desired accuracy for the result had been obtained, and represented an enormous saving in both wallclock and CPU time.

 For examples such as the above, search algorithms based on some metric measure are the alternative to the more ``brute force'' taskfarm. For one-dimensional parameter spaces such search algorithms are relatively straightforward. However, the extension of such algorithms to multidimensional parameter spaces is highly non-trivial. By contrast, steering enables the use of intuition to rapidly guide such parameter space searches.

\section{Conclusions}\label{concs}

The conclusions we derive from the work described above fall into three categories. Firstly, what have we learned about amphiphilic systems by studying them with two different models. Secondly, what have we learned about the relative merits in terms of numerical stability and performance of our two models, and thirdly what lessons may we draw from the performance of this work to enable the ease and efficiency of future high performance computation to be improved.

We have demonstrated previously that our two models can both capture a very wide range of amphiphilic behaviour. In this paper we have drawn attention to those areas where our models produce different results. It is in these areas that studying the same system with different mesoscale models yields useful information. It seems likely from our simulation results that the inability of the lattice gas model to create static structures is due to the small ($1/26$) fraction of rest particles. In our lattice-Boltzmann model, where no such restriction applies, static lamellar and cubic equilibrium phases have been demonstrated to exist. 

Conversely, the absence of noise or stochastic driving forces in our lattice-Boltzmann model means that important modes of domain coarsening are absent. Particularly, Brownian motion effects which arise naturally from the particulate nature of the lattice gas are absent in the model. This leads to the arrest of domain growth for spherical domains which do not have hydrodynamic coarsening mechanisms. The same domains are observed to grow by Brownian collision and coalescence coarsening mechanisms in our lattice-gas model~\cite{bib:maziar1}.

The unconditional numerical stability of our lattice-gas model is a feature which greatly aids computational work with this model. The stability problems of the LBE model outlined above mean that, unlike the lattice-gas model, it currently cannot reproduce the full range of amphiphilic behaviour within one set of model parameters, but requires different sets depending on the mesophase required. The hydrodynamic instabilities ubiquitous to all LBE models which use the BGK approximation may be treated with linear stability analysis, and removed by introduction of a linearised collision operator. As our forcing term is separate from the BGK operator such an approach may yield useful results here, and may enable the model to be stable to higher Reynolds numbers, a matter of some significance in the study of binary immiscible fluids. Treatment of the instabilities produced purely due to the multiphase nature of the model would require a less {\it ad hoc} approach from the outset, however. Recently, lattice-Boltzmann models based on an appropriate Lyapounov functional  with unconditional nonlinear stability have been proposed~\cite{bib:entropicLB}. Such models could unltimately lead to unconditionally stable multiphase models similar to those described here, although this work is at a rather early stage.

Our results regarding the parallel performance of the models clearly demonstrate the large computational advantage obtained by using the lattice-Boltzmann approximation. The almost order-of-magnitude speedup, even before taking into account the obviation of the need to ensemble-average the results, accounts for the massive takeup of the lattice-Boltzmann method in the mesoscale community. Analysis of the relative influence of cache effects (large for small numbers of processors), computational load (dominant) and communication overhead (acceptably small) show that the computation of the (local) fluxes and (nonlocal) fields make the ternary lattice-gas model sensitive to cache effects. In the binary case a modification of the lattice gas model is possible which makes all computations local. Ghost particles are introduced which effectively transmit information about colour gradients during the propagation step. Our model was modified to include such particles, but no significant speedup was seen, indicating that the bottleneck is indeed in the communication to memory, and not the processor-to-processor communication. In addition, no such modification is possible for the ternary code. 

The highly scalable nature of our codes and physical complexity of amphiphilic fluids  means that we are able to produce large quantities of complex data. Such data can only be rapidly understood with the aid of high performance visualisation techniques. However, the use of such techniques on gigabytes of data produced in batch simulations and analysed {\it ex post facto} may take far longer than the simulations themselves. We have found that closely coupling simulation with visualisation through techniques such as computational steering may be done simply and quickly, and can lead to a large reduction in the wallclock time between conceiving a simulation and gaining an understanding of the results of that simulation. 

Our research so far in this area will be extended over the next three years in a UK government (EPSRC) funded e-science testbed called RealityGrid (http://www.realitygrid.org). The mesoscale codes, visualisation and computational steering techniques described in this paper will form the basis of a much more comprehensive and focussed effort to link high performance visualisation, computing and experimental facilities through newly available high bandwidth links.

\section*{Acknowledgements}

We are indebted to numerous people and organisations for their support of this work, including Silicon Graphics Incorporated (particularly Bart van Bloemann Waanders, Daron Green and Rob Jenkins), Advanced Visual Systems (particularly Ian Currington), Oxford Supercomputing Centre, the EPSRC E7 High Performance Computing Consortium (particularly Sujata Krishna), the Edinburgh Parallel Computing Centre (particularly Mario Antonioletti), the National Computational Science Alliance in Illinois, USA,  Boston University Center for Computational Science, CSAR, and Manchester Visualisation Centre (particularly Jo Leng). Resources
for the CSAR T3E and Origin $2000$  were allocated under EPSRC grant number GR/M56234, and for the CSAR O$2000$ by special arrangement with CSAR. Access to a $16$ node SGI Onyx system was provided by the Centre for Computational Science, Queen Mary, University of London.

\bibliographystyle{elsart-num}

\begin{thebibliography}{10}
\expandafter\ifx\csname url\endcsname\relax
  \def\url#1{\texttt{#1}}\fi
\expandafter\ifx\csname urlprefix\endcsname\relax\def\urlprefix{URL }\fi

\bibitem{bib:review1_lga}
D.~H. Rothman, S.~Zaleski, G.~Zanetti, Rev. Mod. Phys. 43 (1991) 4320.

\bibitem{bib:fchc}
U.~Frisch, D.~d'Humieres, B.~Hasslacher, P.~Lallemand, Y.~Pomeau, J.-P. Rivet,
  Complex Systems 1 (1987) 1--31.

\bibitem{bib:lb_review1}
S.~Chen, G.~Doolen, Ann. Rev. Fluid. Mech 30 (1998) 329.

\bibitem{bib:lb_review2}
Y.~H. Qian, S.~Succi, S.~A. Orszag, Annual Reviews of Computational Physics 3
  (1995) 195.

\bibitem{bib:lb_review3}
A.~J.~C. Ladd, Phys. Rev. Lett. 70 (1993) 1339.

\bibitem{bib:lb_review4}
A.~J.~C. Ladd, J. Fluid Mech. 271 (1994) 285.

\bibitem{bib:bgk}
P.~Bhatnagar, E.~P. Gross, M.~K. Krook, Phys. Rev.  (1954) 94--511.

\bibitem{bib:luo2}
L.~S. Luo, Theory of the lattice boltzmann method - lattice boltzmann models
  for non-ideal gases., Phys. Rev. E 62 (2000) 4982.

\bibitem{bib:gelbart}
W.~Gelbart, D.~Roux, A.~Ben-Shaul, Modern Ideas and Problems in Amphiphilic
  Science, Springer, Berlin, 1993.

\bibitem{bib:gs2}
G.~Gompper, M.~Schick, Phase Transitions and Critical Phenomena 16 (1994)
  1--181.

\bibitem{bib:mdmlc}
J.-B. Maillet, V.~Lachet, P.~V. Coveney, Phys. Chem. Chem. Phys. 1 (1999)
  5277--5290.

\bibitem{bib:bce}
B.~M. Boghosian, P.~V. Coveney, A.~N. Emerton, Proc. Roy. Soc. A 452 (1996)
  1221.

\bibitem{bib:bcp}
B.~M. Boghosian, P.~V. Coveney, P.~J. Love, Proc. Roy. Soc. London A. 456
  (2000) 1431.

\bibitem{bib:molsim}
B.~M. Boghosian, P.~V. Coveney, P.~J. Love, J.-B. Maillet, Mesoscale modeling
  of amphiphilic fluid dynamics, Molecular Simulation 26 (2001) 85--100.

\bibitem{bib:LCB}
P.~J. Love, P.~V. Coveney, B.~M. Boghosian, Three dimensional hydrodynamic
  lattice-gas simulations of domain growth and self assembly in binary
  immiscible and ternary amphiphilic fluids., Phys. Rev. E. 64 (2001) 1503.

\bibitem{bib:LCB2}
P.~J. Love, J.-B. Maillet, P.~V. Coveney, Three dimensional hydrodynamic
  lattice-gas simulations of binary immiscible and ternary amphiphilic flow
  through porous media., Phys. Rev. E. 64 (2001) 061302.

\bibitem{bib:LCB3}
P.~J. Love, P.~V. Coveney, Three dimensional hydrodynamic lattice-gas
  simulations of ternary amphiphilic fluids under shear flow., Phil. Trans. R.
  Soc.London Series A In Press.

\bibitem{bib:maziar1}
H.~Chen, B.~M. Boghosian, P.~V. Coveney, M.~Nekovee, Proc. R. Soc. London A.
  456 (2000) 2043.

\bibitem{bib:maziar2}
M.~Nekovee, P.~V. Coveney, H.~Chen, B.~M. Boghosian, Phys. Rev. E 62 (2000)
  8282.

\bibitem{bib:em1}
A.~N. Emerton, P.~V. Coveney, B.~M. Boghosian, Phys. Rev. E 55 (1997) 708.

\bibitem{bib:rk}
D.~H. Rothman, J.~M. Keller, J. Stat. Phys. 52 (1988) 1119--1127.

\bibitem{bib:chli}
C.~K. Chan, N.~Y. Liang, Europhys. Lett. 13 (1990) 495--500.

\bibitem{bib:shanchen1}
X.~Shan, H.~Chen, Phys. Rev. E. 47 (1993) 1815.

\bibitem{bib:shanchen2}
X.~Shan, H.~Chen, Phys. Rev. E. 49 (1994) 2941.

\bibitem{bib:luo1}
L.~S. Luo, Unified theory of lattice boltzmann models for nonideal gases, Phys.
  Rev. Lett. 81 (1998) 1618.

\bibitem{bib:martys}
N.~Martys, X.~Shan, H.~Chen, Phys. Rev. E 58 (1998) 6855.

\bibitem{bib:basleb}
S.~Bastea, J.~L. Lebowitz, Phys. Rev. Lett. 78~(18) (1997) 3499.

\bibitem{bib:basleb2}
S.~Bastea, R.~Esposito, J.~L. Lebowitz, R.~Marra, J. Stat. Phys. 101 (2000)
  1087.

\bibitem{bib:catesJFM}
V.~M. Kendon, M.~E. Cates, I.~Pagonabarraga, J.~C. Desplat, P.~Bladon, Inertial
  effects in three-dimensional spinodal decomposition of a symmetric binary
  fluid mixture: a lattice boltzmann study, Journal of Fluid Mechanics  (2001)
  147--203.

\bibitem{bib:maziar4}
M.~Nekovee, P.~V. Coveney, J. Amer. Chem. Soc. .

\bibitem{bib:wagneryeo}
A.~J. Wagner, J.~M. Yeomans, Phys. Rev. Lett. 80 (1998) 1429--1432.

\bibitem{bib:nelido}
N.~Gonz\'{a}les-Segredo, M.~Foster, in: Proceedings of the Sixth European
  SGI/Cray MPP Workshop, 2000.

\bibitem{bib:MPI}
Mpi: A message passing interface, in: Proc Supercomputing '93, Message Passing
  Interface Forum series, IEEE Computer Society, pp. 878--883.

\bibitem{bib:Measuresetal}
K.~M. Measures, J.~M. Martin, R.~C. McLatchie, Supercomputing resource
  management -- experience with the sgi cray origin 2000, in: B.~Cook (Ed.),
  Architectures, Languages and Techniques for Concurrent Systems, IOS Press,
  1999.

\bibitem{bib:beazlom}
D.~M. Bezley, P.~S. Lomdahl, Lightweight computational steering of very large
  scale molecular dynamics simulations, in: Proceedings of Supercomputing 96,
  1996.

\bibitem{bib:VIPAR1}
S.~Larkin, A.~J. Grant, W.~T. Hewitt, Libraries to support distribution and
  processing of visualisation datasets, Future Generation Computing Systems
  (Special Issue {HPCN} 96) 12~(5) (1996) 431--440.

\bibitem{bib:CUMULVS}
G.~A. {Geist II}, J.~A. Kohl, P.~M. Papadopolous, {CUMULVS}: Providing
  fault-tolerance, visualisation and steering of parallel applications,
  International Journal of High Performance Computing Applications 11~(3)
  (1997) 224--236.

\bibitem{bib:cactus}
{C}actus {D}evelopment {T}eam, \texttt{http://www.cactuscode.org}.

\bibitem{bib:avs}
Advanced Visual Systems, Inc., {AVS} User's Guide, Release 4 (1992).

\bibitem{bib:PARVTK}
J.~Ahrens, K.~Brislawn, K.~Martin, B.~Geveci, C.~C. Law, M.~Papka, Large-scale
  data visualization using parallel data streaming, IEEE Computer graphics and
  applications 21~(4) (2001) 34--41.

\bibitem{bib:entropicLB}
B.~M. Boghosian, J.~Yepez, P.~V. Coveney, A.~Wagner, Proc. R. Soc. (London) A.
  457 (2001) 717.

\end{thebibliography}

\end{document}